\documentclass[fleqn,usenatbib]{mnras}

\usepackage{newtxtext,newtxmath}
\usepackage[T1]{fontenc}
\usepackage{orcidlink}

\usepackage{hyperref}
\usepackage{lipsum}
\usepackage{bm}
\usepackage{threeparttable}

\usepackage{array}
\usepackage[compatibility=false]{caption}
\usepackage{subcaption}         
\usepackage{multirow}

\usepackage{lscape}             
\usepackage{placeins}           
                                
\usepackage{xcolor}
\usepackage{rotating}

\usepackage{graphicx}	
\usepackage{amsmath}	

\newcommand{\red}[1]{\textcolor{black}{#1}}

\newcolumntype{P}[1]{>{\centering\arraybackslash}p{#1}}

\DeclareRobustCommand{\VAN}[3]{#2}
\let\VANthebibliography\thebibliography
\def\thebibliography{\DeclareRobustCommand{\VAN}[3]{##3}\VANthebibliography}

\title[Uncertainty-Aware Tidal Disruption Event Classification]{Uncertainty-Aware Tidal Disruption Event Classification : A Host-Agnostic Probabilistic Random Forest Approach}

\author[V. Anilkumar et al.]{
Vysakh Anilkumar$^{1}$\orcidlink{0009-0008-3146-287X},
Sjoert van Velzen$^{1}$\orcidlink{0000-0002-3859-8074},
Marek Kowalski$^{2}$\orcidlink{0000-0001-8594-8666},
Simeon Reusch$^{3}$\orcidlink{0000-0002-7788-628X}
\\
$^{1}$Leiden Observatory, Leiden University, Postbus 9513, 2300 RA Leiden, The Netherlands\\
$^{2}$Deutsches Elektronen-Synchrotron (DESY), Platanenallee 6, D-15378 Zeuthen, Germany \\
$^{3}$Institut fur Physik, Humboldt-Universit\"{a}t zu Berlin, D-12489 Berlin, Germany
}

\date{Accepted XXX. Received YYY; in original form ZZZ}

\pubyear{2026}

\begin{document}
\label{firstpage}
\pagerange{\pageref{firstpage}--\pageref{lastpage}}
\maketitle

\begin{abstract}
The classification of Tidal Disruption Events in large-scale photometric surveys is challenging because deterministic machine learning models produce overconfident misclassifications under varying data quality and low signal-to-noise conditions. \red{Existing lightcurve-based approaches do not propagate measurement uncertainties properly into the classification itself, consequently generating brittle outputs.} We present a host-agnostic, uncertainty-aware classification framework utilizing a Probabilistic Random Forest (PRF). Our pipeline extracts 11 characteristic features—including rise and decay timescales, blackbody temperature evolution, and pre-transient variability metrics—from nuclear transients in the ZTF alert stream. Relying exclusively on photometric lightcurve data without host galaxy information ensures effectiveness for the faint transient population expected from the Rubin Observatory. The PRF classifier, which treats feature measurements as distributions, was evaluated against XGBoost through a Leave-One-Out Cross-Validation strategy. The analysis demonstrates the PRF yields higher stability and robustness for ambiguous candidates compared to the predictions of XGBoost. The two classifiers occupy complementary regimes: XGBoost achieves higher recall in balanced and high-precision scenarios, where its rigid decision boundaries efficiently isolate TDE-like sources, while PRF rejects more false positives by penalizing sources with large feature uncertainties. Applying this validated framework to archival data identified 11 new candidate TDEs from the unclassified population and isolated 3 potential photometric TDEs within existing training labels previously misclassified as supernovae or active galactic nuclei. This work demonstrates that TDEs can be reliably identified using photometric lightcurve features, providing a \red{host-agnostic} framework. Uncertainty-aware, probabilistic classifiers are essential for the Rubin era to prevent the overconfident misclassifications inherent in deterministic models operating at low signal-to-noise.
\end{abstract}



\section{Introduction}
Tidal Disruption Events (TDEs) occur when a star ventures within the tidal radius of a supermassive black hole (SMBH), where the tidal forces overcome the star's self-gravity, resulting in disruption. Following this event, a fraction of the stellar debris is ejected into interstellar space, while the rest remains gravitationally bound and falls back to (eventually) form an accretion disk \citep{rees_tidal_1988}. This accretion process powers a highly luminous transient flare lasting from months to years, observable across the electromagnetic spectrum \citep{gezariTidalDisruptionEvents2021}. Consequently, TDEs serve as a unique and powerful astrophysical probe for otherwise quiescent SMBHs residing in galactic centers, offering insights into SMBH demographics, accretion disk formation, and astrophysical jet launching mechanisms. 

To date, approximately 150 TDEs have been discovered, predominantly in the optical regime. This expanding catalog is largely driven by large-scale photometric surveys, most notably the Zwicky Transient Facility (ZTF) \citep{bellmZwickyTransientFacility2019, dekanyZwickyTransientFacility2020}, which has identified over 100 TDEs since its inception in 2018 \citep[eg. ][]{vanVelzen21, Hammerstein23, yaoTidalDisruptionEvent2023}. Despite these initial discoveries occurring via photometric transient lightcurves, the definitive classification of TDEs still relies heavily on spectroscopic follow-up. This approach remains sustainable within the operational limits of ZTF, where the detection threshold of roughly 20.5 magnitudes yields candidates bright enough for spectroscopic classification \citep{grahamZwickyTransientFacility2019}. However, time-domain astronomy is approaching a critical transition with the upcoming Vera C. Rubin Observatory (hereafter Rubin) \citep{ivezicLSSTScienceDrivers2019}. Rubin is projected to discover orders of magnitude more TDEs \citep{bricmanProspectsObservingTidal2020}, the vast majority of which will fall far below the spectroscopic follow-up limit \citep{yaoTidalDisruptionEvent2023}. This necessitates a shift away from spectroscopic reliance toward purely photometric classification frameworks.

In recent years, several machine learning (ML) architectures have been proposed to classify TDEs photometrically, primarily trained and tested on ZTF data \citep{gomez_identifying_2023, Stein24, sheng_neural_2024, pavez-herrera_alerce_2025,bhardwajPhotometricClassifierTidal2025, lanzaEarlyIdentificationOptical2026}. However, most existing models exhibit two limitations that hinder their scalability to the Rubin era. First, many frameworks rely heavily on host galaxy features, which introduces significant caveats. Because Rubin will detect TDEs at much higher magnitudes, a significant fraction of host galaxies will lack WISE data or other AGN indicators \citep{westonIdentifyingTransientHosts2025} that are important for TDE identification. \red{Furthermore, the known TDE population is strongly over-represented in E+A/post-starburst hosts} \citep{French2016, vanVelzen21, Hammerstein23, wangExplanationOverrepresentationTidal2024}. Models conditioned on host features therefore risk inheriting and \red{encoding this skewness in the host population} rather than capturing the intrinsic photometric signatures of the TDEs themselves. Consequently, developing host-agnostic classification frameworks that rely exclusively on the transient lightcurve is imperative. \red{Here we define host-agnostic to a classifier for which every input feature is derived solely from the difference flux lightcurve, with no measured property of the host galaxy entering the model. By host properties we mean global properties of the galaxy, such as its brightness and its mid-infrared color, redshift, and star-formation history. While our cut on host-flare separation could be considered as a host galaxy property, we do not use this as a feature as an input to the model. So by design, our classifier is host-agnostic for nuclear transients.}

\red{Given this definition, existing classifiers can be placed relative to it, and several of them already draw the same distinction internally. \citet{sheng_neural_2024} train their model in both a transient-and-host and a transient-only configuration, where the latter (NEEDLE-T) uses no host information. \citet{lanzaEarlyIdentificationOptical2026} similarly train a nuclear classifier that includes the host offset (\texttt{distnr}) as a feature alongside a broad classifier that omits it, motivated by the offset becoming unreliable when the host is undetected in quiescence. \citet{steinTDE2025abcrTidal2026} separate three cases explicitly, with a standard \texttt{tdescore} using lightcurve and host features, an unbiased variant using lightcurve information together with a nuclear selection, and a fully agnostic off-nuclear variant using the lightcurve alone. \citet{bhardwajPhotometricClassifierTidal2025} classify simulated LSST lightcurves from the ELAsTiCC2 challenge using lightcurve features alone, without host associations or redshifts. The framework presented here is host-agnostic in the same sense as the unbiased variant of \texttt{tdescore}, with candidates drawn from a nuclear alert filter and only lightcurve-derived features reaching the classifier. }

\red{The second limitation of these frameworks whether or not they use host information, is that majority of them do not propagate measurement uncertainty into the classification itself.} These models extract a set of morphological features from the lightcurves and subsequently treat them as exact point estimates during classification. \red{\citet{lanzaEarlyIdentificationOptical2026} also use the uncertainty on the fitted reference time as an additional input feature, using it as an indicator of how well the lightcurve fit is constrained. But this acts as a global measure of fit quality, and the uncertainties on the individual features themselves remain outside the decision boundaries. While such approaches are effective on high-cadence, high-signal ZTF data, they break down when applied to the sparser and noisier data expected from fainter candidates in the Rubin era. Without propagating the uncertainty from the feature extraction, and thus the information about the data quality, into the classification, such models produce brittle, overconfident (mis)classifications when processing low signal-to-noise ratio transients. }

In this work, we introduce a machine learning architecture designed to explore the possibility of addressing both the host-association and data-quality bottlenecks simultaneously using only the transient light curve. To resolve host reliance, we implement a feature engineering pipeline that depends exclusively on multi-band photometric properties. By intentionally avoiding host data, we aim to establish a framework that remains effective for the faintest transients and remains unbiased by host environments, ensuring it can theoretically operate on the 10 million alerts per night expected from LSST. To address the data-quality challenge, we utilize a Probabilistic Random Forest \citep{Reis19} as our classifier. This architecture is intended to provide a stable, conservative alternative to deterministic models, with the goal of evaluating how effectively measurement variance can be handled in a sparse, low-signal regime. Crucially, our pipeline is designed for photometric classification using complete light curves, rather than acting as an early warning system which are optimized for real-time follow-up during the rising phase \citep[eg.][]{lanzaEarlyIdentificationOptical2026}.

This paper is structured as follows: Section \ref{sec:data} describes the photometric dataset and spectroscopic labels utilized in this study. Section \ref{sec:Feat_engineer} details the feature engineering process, including multi-band lightcurve modeling and parameter inference methodologies. In Section \ref{sec:ClassFrame}, we define our classification framework and evaluate the performance of the PRF model against a deterministic XGBoost baseline. Section \ref{sec:discussion} discusses the implications of our results and applies the trained classifier to search for new, unclassified TDE candidates within the archival ZTF data stream. Finally, Section \ref{sec:conclusion} summarizes our conclusions and outlines directions for future work.

\section{Photometric Data and Spectroscopic Labels}\label{sec:data}

\subsection{ZTF nuclear transient dataset}\label{subsec:nuc_dataset}
We identify nuclear transient candidates by processing the ZTF alert stream \citep{masci_zwicky_2018} through a dedicated filter within the AMPEL framework \citep{Nordin19}. This filter, originally used by \citet{vanVelzen19, vanVelzen21, Hammerstein23},  broadly selects sources based on photometric reliability, star-galaxy classification, and its proximity with the host nucleus ($< 0.5^{\prime\prime}$). These criteria are designed to be loose, prioritizing high recall over precision. The application of this filter on archival data until mid-2023, resulted in a primary nuclear transient dataset comprising 12976 sources. Full details on the filter's specific cuts can be found in \citet[chap.~6]{Reusch24}.

\subsection{ZTF Lightcurve}\label{subsec:lcdata}
We retrieved the full ZTF forced photometry lightcurves \citep{Masci23}  for every source identified in the alert photometry, spanning approximately six years (March 2018 to September 2024).
We apply preliminary quality cuts described in \citet{Masci23} retaining only epochs with successful execution of forced photometry  (\texttt{procstatus} = 0), good photometric calibration (\texttt{infobitssci} < 33554432), low background noise (\texttt{scisigpix} $\leq$ 25), and acceptable seeing (\texttt{sciinpseeing} < 4.0"). Additionally, we perform stricter checks on photometric stability by rejecting epochs with high zero-point dispersion (\texttt{zpmaginpscirms} < 0.05) or significant deviations from the median zero-point ($\lvert \frac{\log_{10}{ZP}}{\rm{med(ZP)}} \rvert$ < 0.4). If the source is present in more than one fields, the data from the field with most data points is used. Even though ZTF has $i$-band coverage, we only use both the $g$ and $r$ bands for subsequent analysis, as they are primarly provided in a uniform 2 day cadence.  We further remove any source with fewer than 30 datapoints in either of these bands (discarded data contains 2 TDEs- TDE2019cmw and TDE2018hyz). We also remove duplicates in the nuclear dataset, i.e., sources which have been assigned multiple ZTF ids, bringing the final nuclear dataset to 12251 sources. 

We perform galactic extinction correction on these sources using the dustmaps from \citet{SchlaflyFinkbeiner11} along with extinction law from \citet{Fitzpatrick99}.

The final processing step involves performing a baseline correction on each individual lightcurve to correct for any offsets present after difference imaging photometry. The baseline was established by calculating the inverse-variance weighted average of the flux \red{more than 100 days prior to the transient's peak}. \red{The peak of the transient is taken as the brightest epoch of the combined $g$ and $r$ lightcurve, smoothed using a 1D Gaussian filter ($\sigma = 3$ days).} This average value was subtracted from the lightcurve, with its associated uncertainty propagated, thereby centering the mean of the pre-transient flux to zero. \red{This step is applied in the same way to all nuclear transients. For AGN, the peak corresponds to the brightest epoch in the lightcurve, and once baseline corrected the pre-peak flux is centered at zero but retains the variability of the source. This residual variability is itself used as a classification feature (Section \ref{subsec:featextract}).}

\subsection{Spectroscopic labels}\label{subsec:spec_labels}

To prepare the training dataset, we obtained spectroscopic labels for TDEs and Supernovae (SNe). We cross-matched the nuclear transient dataset with the ZTF Bright Transient Survey \citep{Perley20, Fremling20}, the Transient Name Server (TNS)\footnote{\href{https://www.wis-tns.org/}{https://www.wis-tns.org/}}, and the manyTDE repository\footnote{\href{https://github.com/sjoertvv/manyTDE/tree/main}{https://github.com/sjoertvv/manyTDE.git}}\citep{Mummery24}. TDEs found only in the TNS were manually inspected for ambiguity, leading to the exclusion of TDE2020ukj and TDE2019gte. This resulted in a preliminary sample of 64 TDEs and 1042 SNe. Since we are interested in TDEs, we did not use the detailed spectral classifications for the supernovae. For labels for Active Galactic Nuclei (AGN), we cross-matched our dataset with \cite{Liu19}, the WISE AGN catalog \citep{Assef18}, and the milliquas catalog \citep{Flesch2023}, identifying 5113 AGN. Finally, because repeating TDEs are beyond the scope of this work, we removed five such candidates (TDE2022exr, TDE2021mhg, TDE2020acka, AT2018mac, TDE2021uqv), bringing the final count of TDEs in the training dataset to 59. The list of TDEs used in this work is given in Table \ref{tab:tde_list} in the Appendix.

\begin{table}
    \centering
    \begin{tabular}{c|c|c|c|c}
        Class & TDE & SN  & AGN  & Unknown \\
        \hline
        Count & 59  & 1042 & 5113 & 6030 \\
    \end{tabular}
    \caption{Spectroscopic label summary}
    \label{tab:counttable}
\end{table}

\section{Feature Engineering}\label{sec:Feat_engineer}
In this work, we use a feature-based classification approach that relies on extracting characteristic features from multi-band light curves. These extracted features will subsequently be used to distinguish TDEs from other nuclear transients using machine learning algorithms. Given that we are utilizing only the light curves for classification, and not any host galaxy information, it is essential to primarily capture the shape and structure of the transient event, as well as the correlation between observations in different frequency bands. Therefore, we perform multi-band modeling to achieve this. 

\subsection{Multi-band Lightcurve Model}\label{subsec:lc_model}

Our base model for the transient component is adapted from the light curve model presented in \citet{vanVelzen21}, which is characterized by a Gaussian rise to the peak followed by an exponential decay. We further extend this model to account for the persistent, late-time flux plateau that has been observed in most TDEs \citep{vanvelzenLatetimeUVObservations2019, Mummery24}. This plateau component is modeled using a sigmoid function that smoothly rises from zero to the final plateau flux value, $F_{\rm plat}$. The rise of this plateau component is defined to start at the peak of the transient and reach saturation approximately 30 days later. The explicit addition of $F_{\rm plat}$ enables a distinct measurement of the plateau, which could serve as a crucial feature for distinguishing TDEs from other transients.

The total multi-band flux, $F(t, \nu)$, is therefore the sum of the plateau component and the transient component, given by the following equation,

\begin{equation}
    \label{eq:lctime}
    F(t) = \frac{F_{\rm plat}}{1 + 99\,e^{-\frac{t-t_{\rm peak}}{\tau_{\rm plat}}}}  +  F_{\rm peak} \times \begin{cases}
        
                        e^{-\frac{(t-t_{\rm peak})^2}{2\sigma_{\rm rise}^2}} & t \leq t_{\rm peak}\\
                        e^{-\frac{t-t_{\rm peak}}{\tau_{\rm decay}}} & t \geq t_{\rm peak}\\
                        \end{cases}
\end{equation}

\begin{equation}
    \label{eq:lcmodel}
    F(t, \nu) = \frac{B_{ \nu}(T(t))}{B_{ \nu_{0}}(T(t))} \times F(t)
\end{equation}

In the temporal model defined by Eq.~\ref{eq:lctime}, $F_{\rm peak}$ is the peak flux of the light curve, $t_{\rm peak}$ is the time of peak flux, $\sigma_{\rm rise}$ is the rise coefficient, and $\tau_{\rm decay}$ is the decay coefficient. For the plateau component, $F_{\rm plat}$ is the plateau flux, and $\tau_{\rm plat} = 3.264$ days, which is set to ensure the sigmoid reaches \red{from 1\% of $F_{\rm plat}$ at $t_{\rm peak}$ to} 99\% of $F_{\rm plat}$ near $t_{\rm peak} + 30$ days. \red{The factor of 99 in the sigmoid function comes from placing the midpoint of the sigmoid at 15 days from $t_{\rm peak}$, allowing the sigmoid to start rising at the transient peak}. 

To enable simultaneous multi-band fitting, we use a blackbody scaling term, $\frac{B_{ \nu}(T(t))}{B_{ \nu_{0}}(T(t))}$, which scales the model flux by assuming the source radiates as a blackbody. Instead of assuming a constant blackbody temperature, the temperature, $T(t)$, is modeled using a piecewise linear function characterized by three parameters: the peak temperature ($T_{\rm peak}$), the temperature gradient ($dT/dt$), and the plateau temperature ($T_{\rm plat}$). \red{Both $T_{\rm peak}$ and $T_{\rm plat}$ are capped in the range $10^3$ -- $10^6$ K to prevent overflow during parameter inference.} This piecewise function allows for a simple yet flexible model for temperature evolution to accommodate different types of transients.

\begin{equation}
    \label{eq:temp_evol}
    \log_{10}T(t) = \begin{cases}
            \log_{10}T_{\rm peak} & t < t_{\rm peak} \\
            \log_{10}T_{\rm peak} + \frac{d\log_{10}T}{dt} (t - t_{\rm peak}) & t_{\rm peak} \leq t \leq  t_{\rm plat}\\
            \log_{10}T_{\rm plat} & t\geq t_{\rm plat}
            \end{cases}
\end{equation}

where $t_{\rm plat}$ is the estimated time from which the transient component is negligible or the plateau emission is prominent. \red{We note that this is distinct from the rise of the plateau component in Eq. \ref{eq:lctime}, which begins at $t_{\rm peak}$ and saturates in 30 days. $t_{\rm plat}$ marks where the plateau dominates the observed flux, not where the modeled plateau component switches on.} Putting the two parts together, the parameters of the temporal model ($F_{\rm peak}$, $t_{\rm peak}$, $\sigma_{\rm rise}$, $\tau_{\rm decay}$, $F_{\rm plat} $) are estimated by fitting the light curve to a reference band ($\nu_0$), which corresponds to the mean effective frequency between the ZTF $g$- and $r$-filters. Concurrently, the scaling between bands models the temperature evolution parameters ($T_{\rm peak}$, $d(\log T)/dt$, $T_{\rm plat}$).

\subsection{Parameter Inference}\label{subsec:parameter_inference}

To estimate the posteriors for the parameters of our model, we use Markov Chain Monte Carlo (MCMC) sampling. Given that the lightcurves of different transients can exhibit varying shapes and strengths, we first estimate the initial values of the model parameters using a piecewise linear fit to the transient light curve. This initialization step serves a dual purpose: ensuring MCMC convergence and selecting the optimal model complexity. 
This initial step is particularly important for transients observed near the end of the time series, where insufficient data points might otherwise cause the full model fit to fail. Utilizing the initial estimates for the rise and decay coefficients, we first determine the transient start time ($t_{\rm start}$) and the plateau start time ($t_{\rm plat}$). $t_{\rm start}$ and $t_{\rm plat}$ are defined as the times before and after the peak brightness, respectively, when the transient flux is \red{1\%} of the peak value\red{, which follows from Eq. \ref{eq:lctime} as}
\begin{equation}
      \label{eq:tstart_tplat}
      \begin{aligned}
          t_{\rm start} &= \max\left( t_{\rm peak} - 3.035\;\sigma_{\rm rise},\;\; t_{\rm peak} - 200 \right)\\[2pt]
          t_{\rm plat}  &= \min\left( t_{\rm peak} + 4.605\;\tau_{\rm decay},\;\; t_{\rm peak} + 400 \right)
      \end{aligned}
  \end{equation}

\red{The second term in each expression acts as a
safety cutoff, bounding the window for very slowly rising transients and for AGN, where the fitted rise and decay coefficients can become arbitrarily large.} The estimated $t_{\rm plat}$ is used for selecting the optimal model complexity that best represents the transient event. If there are \red{fewer than five data points} available in the plateau phase ($t > t_{\rm plat}$), we freeze the plateau flux ($F_{\rm plat}$) to zero, effectively simplifying the model to only fit the rise and decay of the transient. For this simpler, no-plateau model, the temperature evolution is also simplified, containing only $T_{\rm peak}$ and the gradient $dT/dt$, as the linear temperature evolution no longer transitions to a constant temperature ($T_{\rm plat}$).

\begin{figure}
    \centering
    \includegraphics[width=\linewidth, alt={Light curve of TDE2021nwa in r and g bands with fitted model peaking at ~165 uJy, plus lower panel of declining fitted temperature.}]{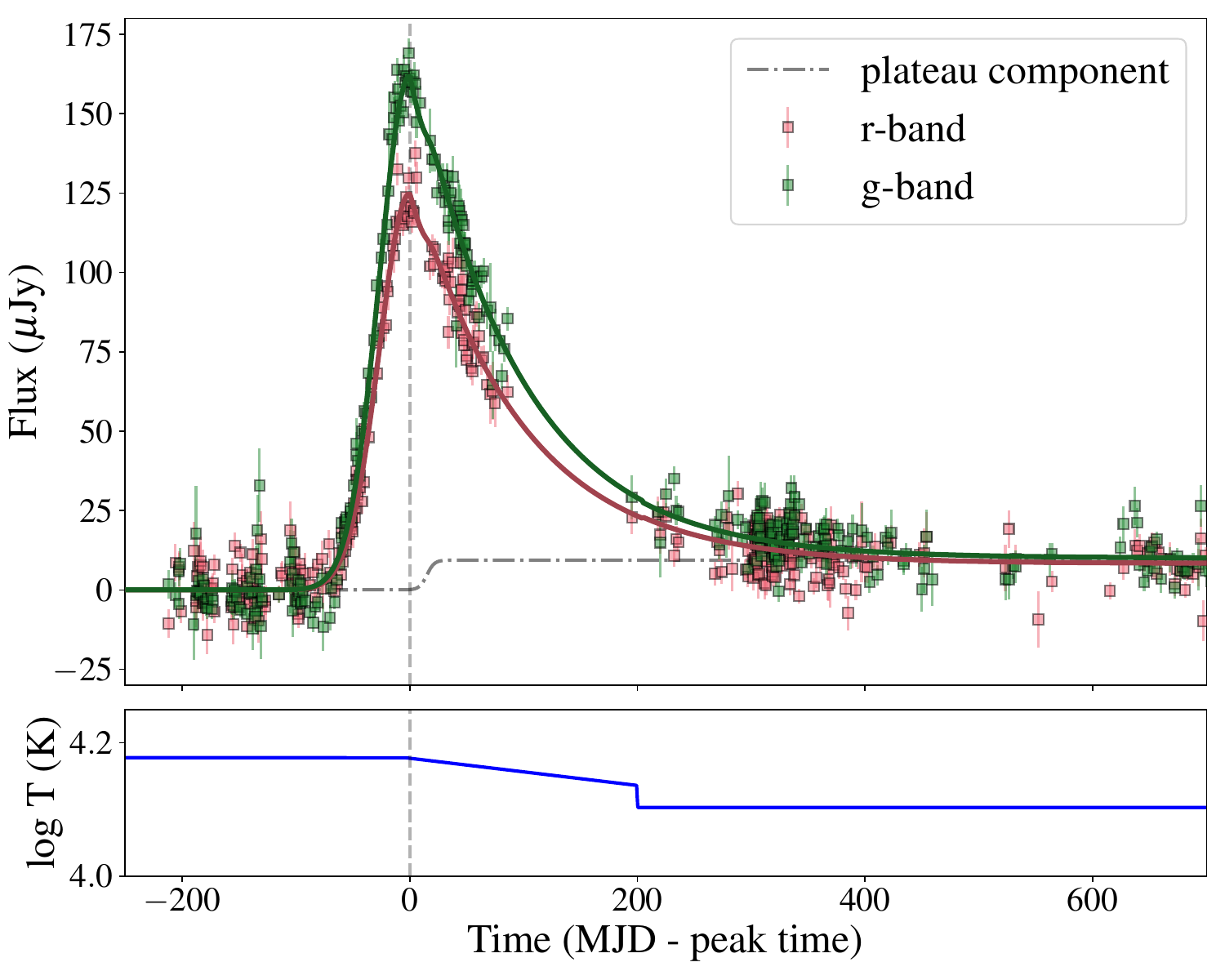}
    \caption{The top panel shows the multi-band lightcurve model fitted to $r$ and $g$ bands of the source TDE2021nwa. The black dot-dash \red{line} shows the component that models the plateau of the TDE. The bottom panel shows the estimated evolution of temperature using the model in equation \ref{eq:temp_evol} \red{, where the abrupt change in slope marks $t_{\rm plat}$, beyond which the temperature is held constant at $T_{\rm plat}$}. }
    \label{fig:tde_fit}
\end{figure}

Once the initial parameters are obtained and the best version of the model is selected, we proceed with a two-stage MCMC sampling for the posteriors.  The first level MCMC focuses exclusively on the late-time data ($\geq t_{\rm plat}$), fitting a constant flux function across all bands, scaled by the blackbody relation. This yields posterior distributions for $T_{\rm plat}$ and $F_{\rm plat}$ using uniform priors. The second level MCMC then fits the full light curve using the complete model (Eq. \ref{eq:lcmodel}). The $T_{\rm plat}$ is fixed to the median value from the level 1 posterior, and the $F_{\rm plat}$ is constrained by a prior based on the kernel density estimation of the level 1 posterior for the same parameter. All other transient and early-time temperature parameters are fitted using broad uniform priors. While fitting the full lightcurve in level 2, for the early-time fit ($\leq t_{\rm plat}$), the temperature evolution $\log_{10}T(t)$ is modeled by a piecewise function (Eq. \ref{eq:temp_evol}) that is constant before $t_{\rm peak}$ and evolves linearly with the gradient $dT/dt$ until the start of the plateau, ensuring a smooth transition to the fixed $T_{\rm plat}$.  An example fit produced by the two-stage MCMC sampling process is shown in Figure \ref{fig:tde_fit}.

We use the \texttt{emcee} sampler \citep{Foreman-Mackey13} for MCMC sampling, employing a Gaussian likelihood function. The final posterior distributions are obtained using 50 walkers for 5000 steps, with the first 3000 steps discarded as burn-in to ensure proper chain convergence. There were a few sources, where the walkers did not converge due to poor data quality (581 unclassified sources and 570 AGN and 13 SNe) and sources which only showed a rising trend in the 7 years of data (82 unclassified sources and 182 AGN). These sources were dropped from the dataset, bringing our total number of sources to  10813.

\subsection{Feature extraction}\label{subsec:featextract}

\begin{table*}
    \renewcommand{\arraystretch}{1.25}
    \caption {Feature summary}
    \label{tab:featsum}
    \centering
    \begin{tabular}{lcp{9cm}}
    \hline\hline
    \textbf{Feature name} & & \textbf{Description}\\
    \hline
    Rise timescale & $t_{\rm rise}$ & The time duration from the start of the transient event (1\% of the $F_{\rm peak}$, measured from the gaussian rise term) until the time of the transient peak. \\
    Decay timescale & $t_{\rm decay}$ & The time duration from the peak flux ($F_{\rm peak}$) until the flux decays to 1\% of the peak flux (measured from exponential decay term)\\
    Peak temperature & $T_{\rm peak}$ & The blackbody temperature estimated at the transient peak derived from multi-band lightcurve modeling\\
    Temperature gradient & $d(\log_{10}T)/dt$ & The rate of change of the log of blackbody temperature over time ($t_{\rm peak} \leq t \leq  t_{\rm plat}$)\\
    Relative plateau strength & $r_{\rm plat}$ & The relative strength of the plateau flux compared to the peak flux ($F_{\rm peak}$)\\
    Plateau SNR & ${\rm SNR_{plat}}$ & The signal-to-noise ratio of the measured plateau emission\\
    Magnitude excess in $g$-band & $\Delta m_g$ & The difference between the peak magnitude of the transient and the ZTF reference magnitude in the $g$-band\\
    pre-transient fractional rms ($g$-band) & $\rm f_{var, g}^{pre}$ & \red{level of variability of the pre-transient 
    flux at the position of the transient}, normalized by the peak flux of the transient ($F_{\rm peak}$) in the $g$-band\\
    pre-transient fractional rms ($r$-band) & $\rm f_{var, r}^{pre}$ & \red{level of variability of the pre-transient 
    flux at the position of the transient}, normalized by the peak flux of the transient ($F_{\rm peak}$) in the $r$-band\\
    plateau fractional rms ($g$-band) & $\rm f_{var, g}^{plat}$ & intrinsic level of variability during the plateau phase, normalized by the peak flux of the transient ($F_{\rm peak}$) in the $g$-band\\
    plateau fractional rms ($r$-band) & $\rm f_{var, r}^{plat}$ & intrinsic level of variability during the plateau phase, normalized by the peak flux of the transient ($F_{\rm peak}$) in the $r$-band\\
    \hline \hline
    \end{tabular}
\end{table*}

Across the literature, several characteristic properties have been identified that help distinguish TDEs from other nuclear transients. Examples of such features include the rise time, the decay time, and the temperature evolution (color evolution). To ensure our features remain relative, we avoid using the peak brightness as a feature, thereby preventing the introduction of a bias toward classifying bright events. In addition to these established properties, we introduce a few new features which show promise in differentiating TDEs from other source classes. These are discussed below.

An important feature that could help separate AGN variability from TDE or Supernova flares is the pre-transient fractional $\mathrm{rms}$: $f_{\rm var, g}^{\rm pre}$. This is calculated using the excess variance ($\sigma_{\rm XS}$) \red{of the pre-transient flux at the position of the transient} ($F_{\rm pre}$), where,

\begin{equation}
    \sigma_{\rm XS}^2 = {\rm Var}(F_{\rm pre}) - \frac{1}{N-1}\sum_{i=1}^{N} \sigma^2_{{\rm err}, i}
\end{equation}

where $N$ is the number of data points and $\sigma^2_{{\rm err}, i}$ is the square of the error for the $i$-th measurement. The pre-transient fractional $\mathrm{rms}$ is then given by,
\begin{equation}
    f_{\rm var, \nu}^{\rm pre} = \frac{\sqrt{\sigma_{\rm XS}^2}}{F_{\rm peak}} 
\end{equation}

\red{This feature measures the level of variability present at the source position before the transient, normalized by the peak flux of the transient ($F_{\rm peak}$). It is computed entirely from the lightcurve and uses no catalogued property of the host. In practice this pre-transient variability is consistent with the noise from the underlying galaxy, and it is this physical interpretation that gives the feature its separating power.} For typical AGN variability, the excess variance for the estimated \red{pre-transient} flux would be of a similar magnitude compared to the estimated peak flux. However, for a strong transient event such as a TDE or a supernova, this ratio would be much smaller because the underlying transient phenomenon is very strong, and the excess variance would be much smaller compared to the peak flux.

We also introduce two features related to the late-time plateau phase of the transient. The first is the relative plateau strength ($r_{\rm plat}$). This feature quantifies the strength of the plateau flux relative to the peak flux ($F_{\rm peak}$). Since TDEs are observed to show plateau emission after the transient event, this feature helps to distinguish sources with a clear measure of a plateau as TDEs. The other plateau feature is the signal-to-noise ratio of the plateau (${\rm SNR_{plat}}$). While $r_{\rm plat}$ measures the relative strength of the plateau, this feature measures its statistical significance. When combined, these two features can serve as a strong pair for flagging TDEs if a plateau is present. 

The full summary of all the 11 features used is given in Table $\rm \ref{tab:featsum}$. Since the machine learning model in this work also requires the uncertainties in these features, uncertainties are calculated for most of them through error propagation methods. For ${\rm SNR_{plat}}$, the uncertainty is set to be $0.0$. For cases where there is only rise and decay measures (i.e., no or very few data points in the plateau phase), any parameters corresponding to the plateau is set to have infinite uncertainty, which corresponds to a missing value as described in Section \ref{subsec:PRF}.

\section{Classification Framework and Model Performance}\label{sec:ClassFrame}
Since we employ a feature-based classification framework, we  utilize tree-based algorithms, which are widely recognized for their higher performance on structured data. Beyond performance, these algorithms offer the distinct advantage of interpretability, which, when coupled with tools such as SHAP values \citep[SHapley Additive exPlanations, ][]{lundberg_unified_2017}, provides transparency into the decision-making process. This explainability is particularly critical when identifying rare classes like TDEs, ensuring that classifications are driven by relevant physical features rather than statistical artifacts. In this work, our primary classifier is the Probabilistic Random Forest (PRF). Additionally, we train models using the XGBoost classifier, a standard in transient classification literature, to provide a baseline for comparing performance metrics. Both classifiers are detailed below.

\subsection{Probabilistic Random Forest Classifier}\label{subsec:PRF}
\begin{figure*}
    \centering
    \includegraphics[width=0.85\linewidth, alt={PRF builds trees in parallel with branching probabilistic paths; XGBoost builds them sequentially on residuals with single deterministic paths.}]{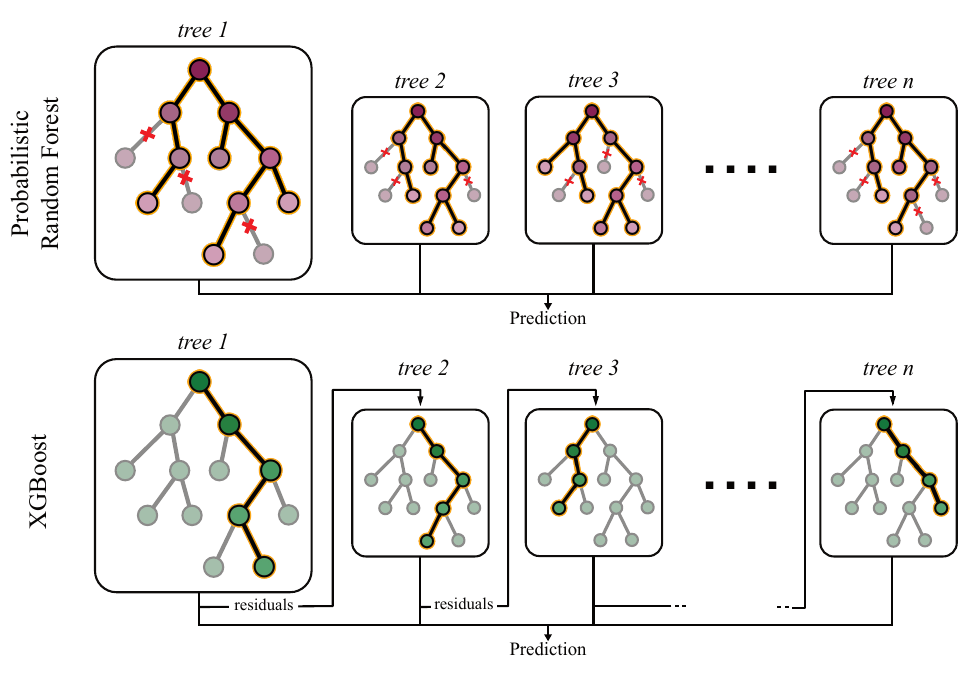}
    \caption{The top panel demonstrates PRF constructing independent trees simultaneously, using probabilistic propagation to route data points across multiple branches using measurement uncertainties at each nodes. Crossed-out paths indicate branches where the propagation probability falls below the defined prediction threshold. The lower panel illustrates XGBoost constructing trees sequentially, relying on deterministic routing to strictly direct instances while training each successive tree on the residual errors generated by preceding structures to compute a cumulative prediction. Throughout both panels, highlighted paths trace the propagation trajectory of a single data object through the respective tree architectures.}
    \label{fig:PRF_XGB_illustration}
\end{figure*}

Introduced by \citet{Reis19}, the Probabilistic Random Forest (PRF) is a modification of the standard Random Forest (RF) algorithm designed to explicitly handle measurement uncertainties. While conventional decision tree algorithms typically treat a feature measurement as a deterministic point assuming infinite precision in its measurement, PRF treats it as a probability density function. This framework allows the model to incorporate the information contained in the error bars directly into the learning process. We are particularly interested in this ability to handle uncertainty in the features, which helps in propagating the uncertainty in feature extraction arising from the model choice and the quality of the data. The top panel of figure \ref{fig:PRF_XGB_illustration} illustrates the propagation of a single data object in the PRF architecture.

The fundamental difference between common decision tree algorithms and PRF lies in how decisions are made at the decision tree nodes. In a standard RF, a node applies a hard threshold to a feature, sending an object exclusively to the right or the left branch. In contrast,  the PRF first calculates the probability that an object satisfies the split condition (using a threshold $\chi_1$ ) given the feature's measurement uncertainty. Consequently, the object propagates to both child nodes with weights proportional to these probabilities. For the $i$-th object, the $k$-th feature and $\chi_1$ threshold, these probabilities are given by,
\begin{equation}
    \pi_i({r}) = F_{i,k}(\chi_1) 
\end{equation} 
\begin{equation}
    \pi_i({l}) = 1 - F_{i,k}(\chi_1) 
\end{equation}

where $F_{i,k}$ is the cumulative distribution. In an ideal PRF, this branching would continue for every object, hence the probability to reach a given node ($n$) from the top node, is given by,
\begin{equation}
    \pi_i(n) = \prod_{\eta \in R} F_{i, k_{\eta}}(\chi_{\eta}) \times \prod_{\xi \in L}{(1 - F_{i, k_{\xi}}(\chi_\xi))}
\end{equation}

where, $R$ and $L$ are the set of right and left turns taken. However, for efficiency, \citet{Reis19} implemented a probability threshold to prune branches with probability less than a given threshold. When it comes to optimization and training, a modified Gini impurity is used. Similar to gini impurity in a random forest, this modification accounts for the probabilistic nature by using expectancy value. For a $m$-class classification problem, the modified gini impurity for node $n$ is given by  

\begin{equation}
    G_n = 1 - \sum^m_{i=0} P^2_{n,A_i}
\end{equation}
where the weighted probability  $P_{n,A_i}$ for class $A_i$ is given by,

\begin{equation}
    P_{n,A_i} = \frac{\sum_{j\in n} \pi(n) \cdot p_{j, A_i} }{\sum_{j\in n} \pi(n)}
\end{equation}

This architecture offers distinct advantages over standard decision trees. PRF handles missing values naturally by treating the measurement as having infinite uncertainty rather than relying on imputation. This makes the model more reliable, particularly when sparse lightcurves result in missing features. Additionally, the algorithm accounts for heteroscedasticity, where feature reliability varies significantly between sources due to differences in brightness and sampling. By using the uncertainties, PRF prioritizes precise measurements while effectively down-weighting those derived from poor quality data. This probabilistic framework also generalizes well when trained on high-quality simulations and applied to real data. It adapts to larger uncertainties by broadening prediction distributions, thereby reducing the risk of confident misclassifications common in deterministic models.

PRF has been applied to various astronomical problems involving variable data reliability, including open cluster membership \citep{PRFcite1}, quasar candidate selection \citep{PRFcite2}, and the identification of young stellar objects \citep{PRFcite4}. More recent studies have further utilized the method for spectral classification \citep{PRFcite3} and the analysis of extragalactic and dusty stellar sources \citep{PRFcite5, PRFcite6}. Most studies indicate that the probabilistic treatment of feature errors provides a robust advantage in low signal-to-noise regimes.

To control overfitting given the limited dataset size, we adopt the hyperparameters suggested for the PRF model by \cite{Reis19}. Specifically, we set the number of estimators to 100 and the maximum depth to 8. The \texttt{keep\_proba} parameter (probability threshold) is set to 0.05, as reducing this threshold further does not yield significant performance improvements \citep{Reis19}. 

\subsection{XGBoost Classifier}\label{subsec:XGB}
Extreme Gradient Boosting (XGBoost) classifier \citep{Chen16} is a scalable implementation of the gradient boosting framework, which differs fundamentally from the bagging approach (bootstrap aggregating, where independent trees are trained in parallel on randomly sampled subsets of the training data) used in Random Forests. While Random Forests construct an ensemble of independent trees in parallel and average their predictions, XGBoost builds trees sequentially. Each subsequent tree is trained to predict the residual errors of the previous ensemble, effectively minimizing a regularized objective function via gradient descent. This sequential correction allows the model to capture complex, non-linear patterns efficiently, making it a standard tool in time-domain astronomy for transient classification \citep[see e.g.][]{Stein24, bhardwajPhotometricClassifierTidal2025, lanzaEarlyIdentificationOptical2026}.

The important distinction between the two classifiers lies in their handling of input data. Unlike the PRF, which explicitly incorporates measurement error distributions into the node splitting logic, XGBoost treats feature measurements as deterministic point estimates. While the algorithm includes mechanisms to handle missing values by learning default directions for optimal splits, it does not natively propagate the uncertainties associated with individual feature measurements. Therefore, we utilize XGBoost primarily as a deterministic baseline. Comparing its performance with the PRF allows us to quantify the specific advantages gained by integrating measurement uncertainties into the classification framework.

\subsection{Model training and validation}\label{subsec:ModelValidation}

With the training data prepared and the classifiers defined, we train the machine learning models using a multi-class framework. Instead of the standard binary approach (TDE vs. non-TDE), we classify events as TDE, SN, or AGN. We prefer this multi-class approach to avoid grouping physically distinct non-TDE sources into a single negative class. Since supernovae and AGN exhibit contrasting variability, combining them creates high intra-class variance, which leads to a complex decision boundary. By modeling these classes separately, the classifier can resolve TDEs against the specific behaviors of stochastic AGN flares and supernovae independently. This structure also enhances interpretability. Maintaining distinct labels allows us to understand feature importance relative to specific physical contaminants, identifying exactly which features distinguish TDEs from each background class. 

\subsubsection{Dealing with data imbalance}\label{subsubsec:dataImbalance}
As reflected in Table \ref{tab:counttable}, there is a severe imbalance in the training sample, where TDEs only represent $\approx$1\% of the training sample. This disparity could significantly affect the model's ability to correctly classify the minority class, as standard classifiers tend to statistically bias the decision boundary toward the majority class to minimize global error. To account for this, we employ an uncertainty-aware extension of the Synthetic Minority Over-sampling TEchnique (SMOTE) algorithm \citep{Chawla11} once features are estimated for individual light curves. This method interpolates in both feature and feature uncertainty space to generate pseudo sources, ensuring that synthetic samples retain consistent observational noise properties. We oversample both the TDE and SN classes to match the population of the AGN class. While generating such a high volume of synthetic data is aggressive, it is necessary to force the classifier to pay attention to the minority class. By densely populating the region around these limited samples, we prevent the model from dismissing them as noise, ensuring it defines a clear decision boundary around TDEs despite their scarcity. While this aggressive oversampling ensures the classifier resolves the minority class, the reliance on linear interpolation in feature space is not ideal, and a more principled approach using physically motivated lightcurve injections or generative models conditioned on TDE parameters is left to future work.

\subsubsection{Leave-One-Out Cross Validation}\label{subsubsec:LOOCV}
\begin{figure*}
    \centering
    \begin{subfigure}{0.41\textwidth} 
        \centering
        \includegraphics[width=0.90\textheight, angle=90, alt={Violin plot of PRF class probabilities for 59 TDEs; blue TDE violins are narrow and near probability 1 for most objects.}]{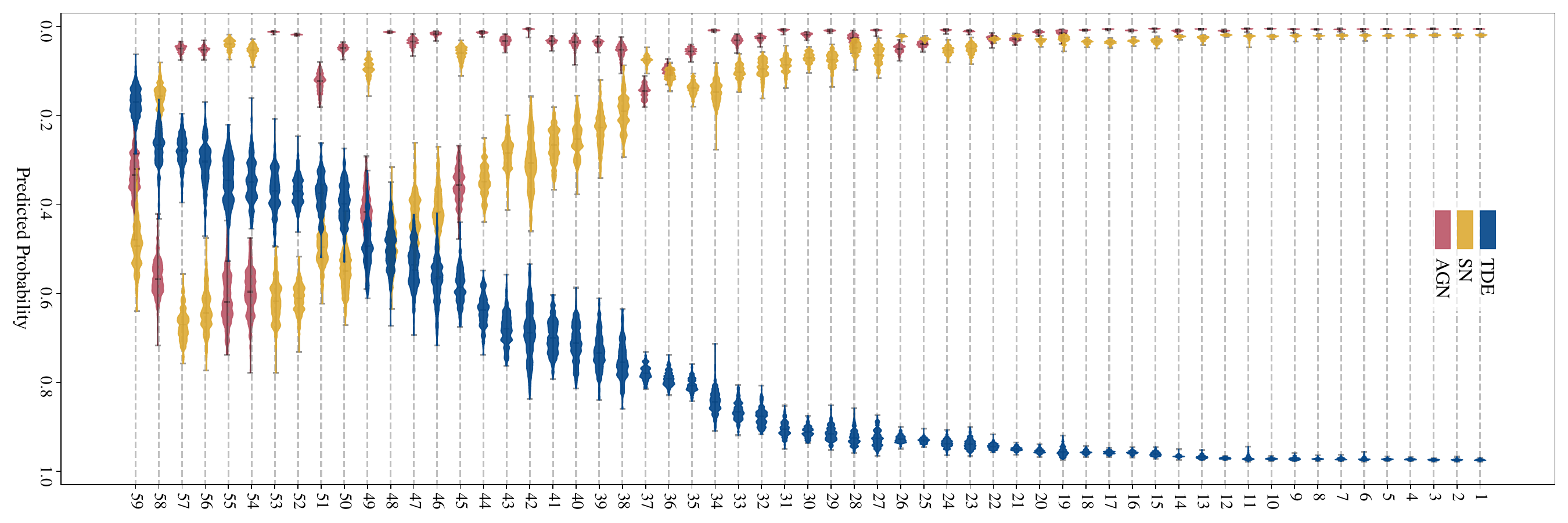}
        \caption{PRF classifier}
        \label{fig:PRFviolin}
    \end{subfigure}
    \begin{subfigure}{0.41\textwidth}
        \centering
        \includegraphics[width=0.90\textheight, angle=90, alt={Violin plot of XGBoost class probabilities for 59 TDEs; distributions are broad and often bimodal, showing unstable predictions.}]{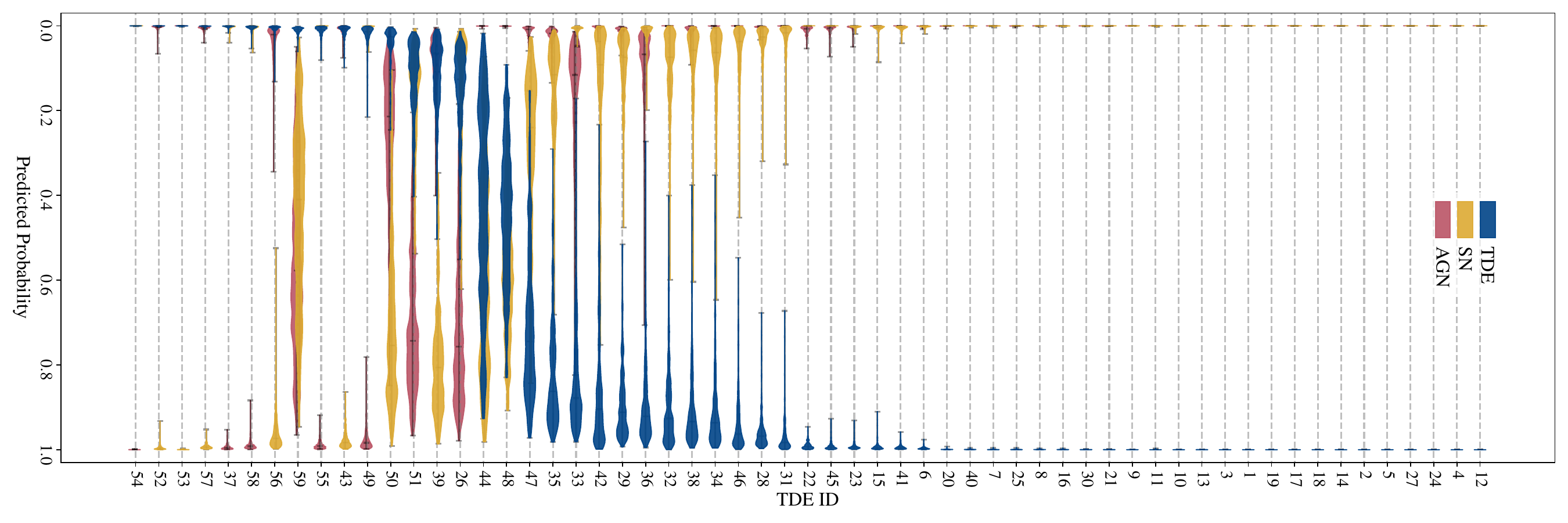}
        \caption{XGB classifier}
        \label{fig:XGBviolin}
    \end{subfigure}
    \caption{This figure shows the results of the LOOCV using violin plots for PRF classifier (left panel) and XGBoost classifier (right panel). The y-axis for both plots display the predicted probability by the corresponding ML algorithm. The y-axis are the TDE IDs assigned to each of the 59 TDEs used in this work. The corresponding ZTF IDs and the TNS names are show in Table \ref{tab:tde_list}. }
    \label{fig:violinplots}
\end{figure*}

Given the scarcity of training samples, a traditional train-validation-test split is unfeasible. Instead, we evaluate model performance using a modified Leave-One-Out Cross-Validation (LOOCV) strategy. LOOCV is a special case of K-fold cross-validation where $K$ equals the number of available sources; here, in our modified version, we set $K=59$, corresponding to the number of TDEs.  In this framework, the dataset is divided into 59 stratified folds, each fold containing exactly one TDE, alongside a selection of SNe and AGN proportional to the class imbalance in the full dataset (approximately 1:16:83). Because the total counts of SNe and AGN are not exact multiples of 59, the distribution of non-TDE sources varies slightly between folds. During each iteration, one fold serves as the test set, while the remaining folds merge to form the training set. Within each iteration, we apply SMOTE to the training set to address class imbalance before training the model as described previously. The SMOTE algorithm is only applied in the training set to prevent data leakage during the validation process. This procedure yields 59 distinct models, providing a prediction for every TDE. To quantify model stability, we repeat the entire cross-validation process 250 times using different random seeds to shuffle the dataset, thereby generating different groups of SNe and AGN in the training set for each iteration.  Grouping these iterations results in a probability distribution for each TDE prediction, offering insight into both the accuracy and stability of the classifier. We visualize these distributions using violin plots, which allow us to visually distinguish between robust classifications, characterized by narrow, consistent probability spreads, and brittle ones that exhibit wide variance depending on the training set. The resulting probability distributions for both PRF and XGBoost are shown in Figure \ref{fig:violinplots}.

Figure \ref{fig:violinplots} demonstrates that the PRF offers higher stability, particularly for difficult candidates. For TDE candidates (e.g., IDs 50-59) that cannot be classified with high confidence, the PRF returns consistently broader probability distributions centered at lower values. This indicates that the model reliably identifies these sources as ambiguous across different training repetitions. In contrast, XGBoost predictions for these same sources are erratic and highly sensitive to the training set composition. As shown in Figure \ref{fig:XGBviolin}, the XGBoost model frequently fluctuates between confident TDE and confident non-TDE classifications depending on the specific realization. This behavior reveals a binary decision boundary where slight variations in the training data trigger significant jumps in predicted probability. Crucially, even for these ambiguous, low-confidence cases, the PRF does not completely rule out the TDE classification. This difference in stability stems primarily from the PRF's bagging architecture,  which averages predictions across independent trees and smoothens decision boundaries through  native uncertainty propagation, in contrast to XGBoost's sequential training which is highly  sensitive to training set composition. This comparison between the two classifiers is further discussed in Section \ref{subsec:PRFvsXGB}.

\subsubsection{Performance metrics}\label{subsubsec:Metrics}

To evaluate classifier performance, we adopt a One-vs-Rest framework where TDEs are treated as the positive class, while Supernovae and AGN are aggregated into a single negative background. Given the significant class imbalance, standard metrics such as accuracy are insufficient, as a trivial model could achieve high accuracy simply by never predicting a TDE. Instead, we report Recall/Completeness, defined as the fraction of true TDEs correctly identified, and Precision/Purity, the fraction of predicted candidates that are true TDEs. To provide a balanced measure between these two metrics, we calculate the F1 score, which is the harmonic mean of precision and recall. Furthermore, we also assess the False Positive Rate (FPR), measuring the fraction of non-TDEs, that were falsely classified as TDEs.

To estimate these metrics, we aggregate the results from the repeated cross-validation by treating each of the 250 repetitions as an independent experiment. For a single repetition, we combine the test predictions from all 59 stratified folds to reconstruct the full dataset. From this pooled set of predictions, we calculate a global confusion matrix and derive the corresponding Precision, Recall, FPR, and F1 scores for a given prediction threshold.  This procedure is repeated for each repetition, generating a distribution for each metric. We report the final performance as the median of these distributions, with the 16th and 84th percentiles representing the  uncertainty interval. 

\begin{table*}
    \renewcommand{\arraystretch}{1.5}
    \centering
    \begin{tabular}{c|m{6em}|P{6em}|P{6em}|P{6em}}
    \hline \hline
       \multirow{2}{6em}{Classifier}  & \multirow{2}{6em}{Metrics} & \multicolumn{3}{c}{Scenarios}  \\ \cline{3-5}
         & & Balanced & High precision & High Recall \\
         \hline \hline
             \multirow{5}{6em}{Probabilistic Random Forest}    & Recall & $0.492^{+0.017}_{-0.017}$  & $0.339^{+0.034}_{-0.017}$  & $0.966^{+0.017}_{+0.000}$  \\
         & Precision & $0.800^{+0.029}_{-0.031}$  & $0.864^{+0.012}_{-0.030}$  & $0.286^{+0.007}_{-0.005}$  \\
         & F1 score & $0.617^{+0.015}_{-0.024}$  & $0.494^{+0.024}_{-0.019}$  & $0.443^{+0.009}_{-0.009}$  \\
         & FPR & $0.001^{+0.000}_{-0.000}$  & $0.001^{+0.000}_{+0.000}$  & $0.027^{+0.001}_{-0.001}$  \\
         \hline
         \multirow{5}{6em}{XGBoost}   & Recall & $0.475^{+0.034}_{-0.017}$  & $0.441^{+0.017}_{-0.034}$  & $0.932^{+0.017}_{+0.000}$  \\
         & Precision & $0.800^{+0.029}_{-0.035}$  & $0.853^{+0.022}_{-0.040}$  & $0.201^{+0.006}_{-0.005}$  \\
         & F1 score & $0.600^{+0.018}_{-0.023}$  & $0.575^{+0.019}_{-0.023}$  & $0.331^{+0.008}_{-0.008}$  \\
         & FPR & $0.001^{+0.000}_{-0.000}$  & $0.001^{+0.000}_{+0.000}$  & $0.041^{+0.001}_{-0.001}$  \\
         \hline \hline
         
    \end{tabular}
    \caption{Metric summary}
    \label{tab:metrics}
\end{table*}

\begin{figure*}
    \centering
    \begin{subfigure}{0.46\textwidth}
        \centering  
        \includegraphics[width=\textwidth, alt={Precision, recall and F1 versus probability threshold for PRF; F1 plateaus near 0.62 and the curves cross around threshold 0.8.}]{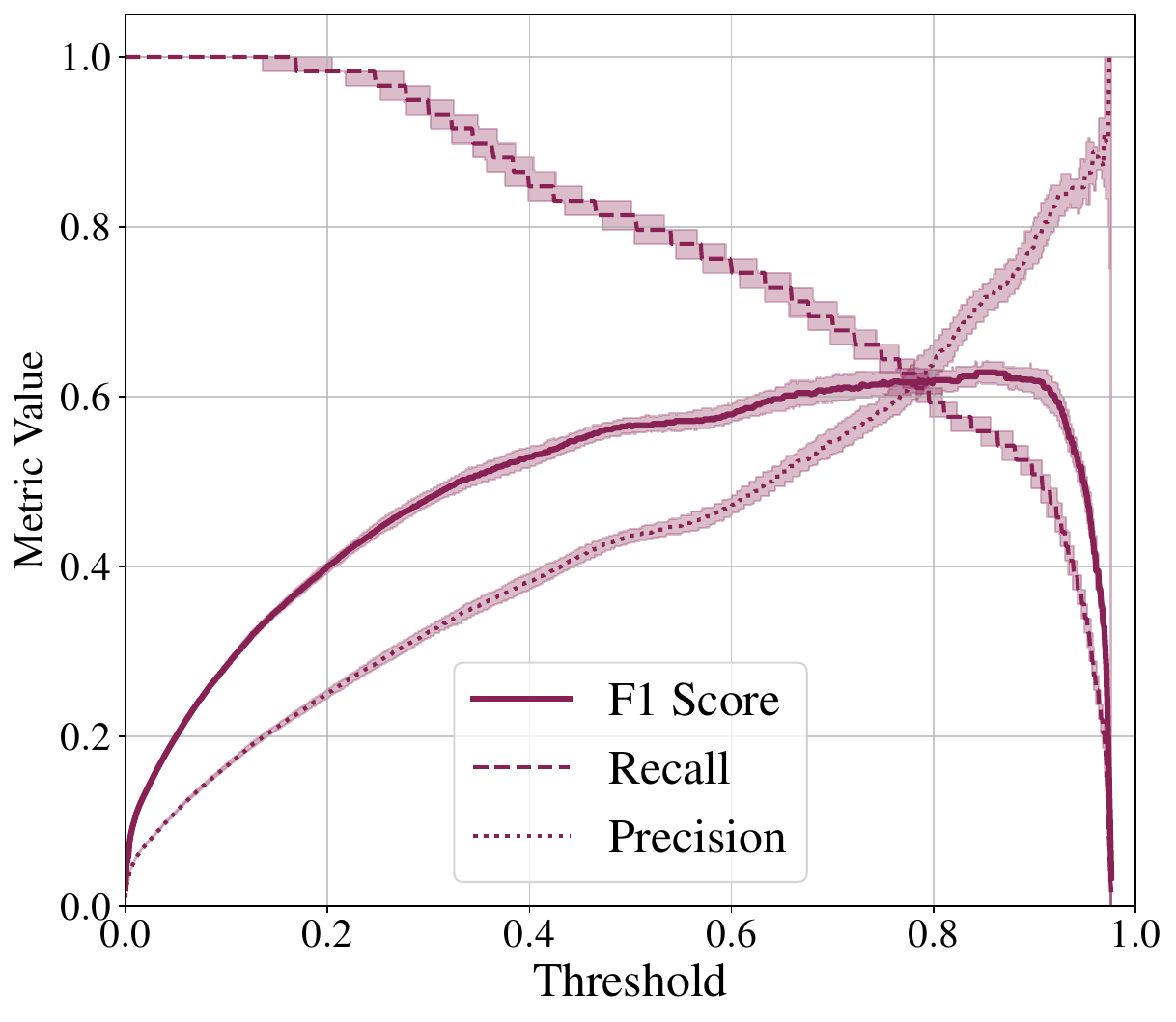}
        \caption{PRF classifier}
        \label{fig:threshPRF}
    \end{subfigure}
    \begin{subfigure}{0.46\textwidth}
        \centering  
        \includegraphics[width=\textwidth, alt={Precision, recall and F1 versus probability threshold for XGBoost; all three stay nearly flat near 0.65 across the mid-range.}]{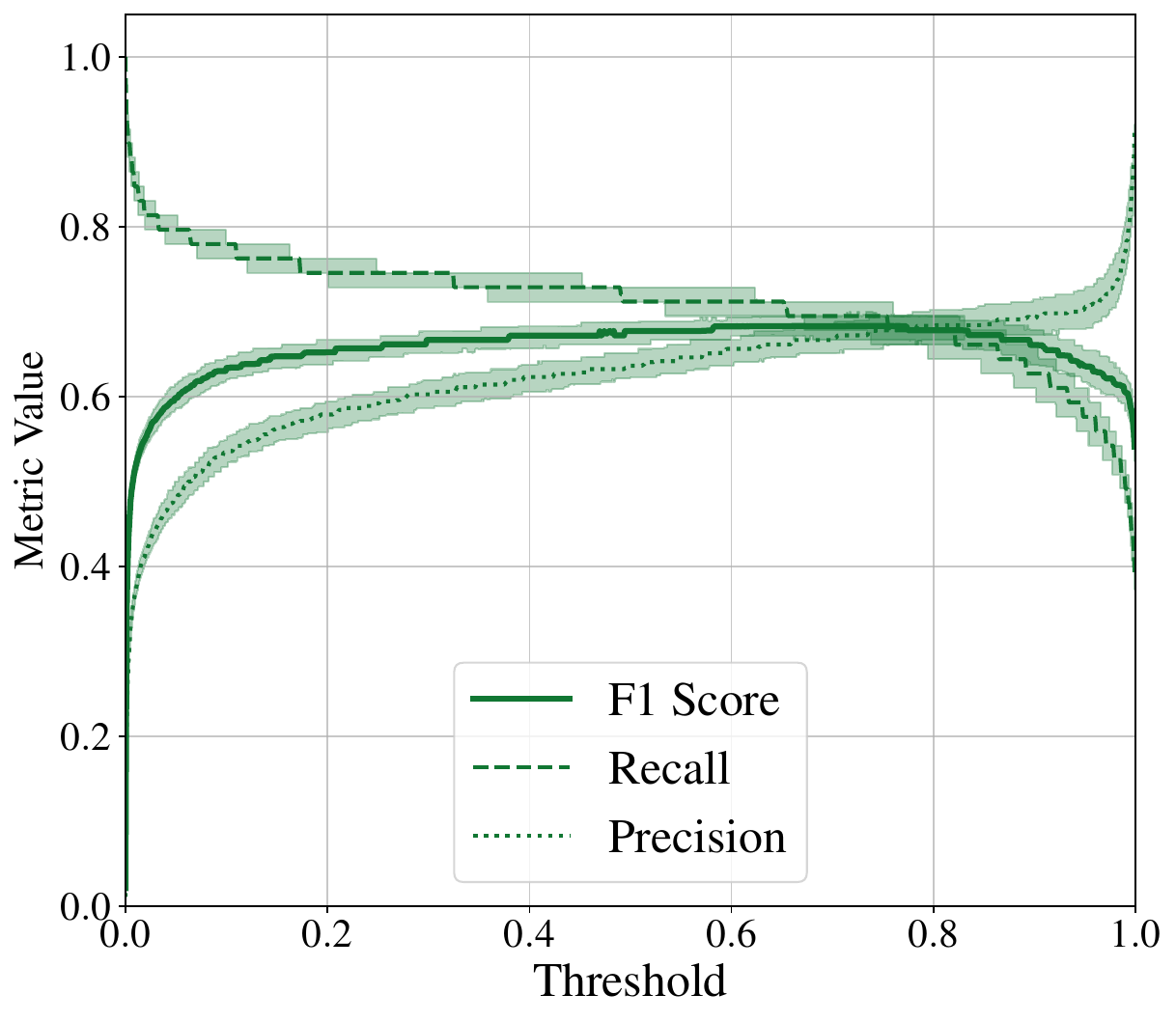}
        \caption{XGB classifier}
        \label{fig:threshXGB}
    \end{subfigure}
    \caption{This figure displays the threshold tuning results for the PRF (left panel) and XGBoost (right panel) classifiers. The plotted lines represent the median Precision, Recall, and F1 Score as a function of the probability threshold across the 250 LOOCV iterations. The shaded regions denote the 16th and 84th percentile uncertainty intervals for each evaluated metric.}
    \label{fig:thresh}
\end{figure*}

\begin{figure*}
    \centering
    \begin{subfigure}{0.36\textwidth} 
        \centering
        \includegraphics[width=\textwidth, alt={Three PRF confusion matrices for the balanced, high-precision and high-recall thresholds, with 29, 20 and 57 of 59 TDEs recovered.}]{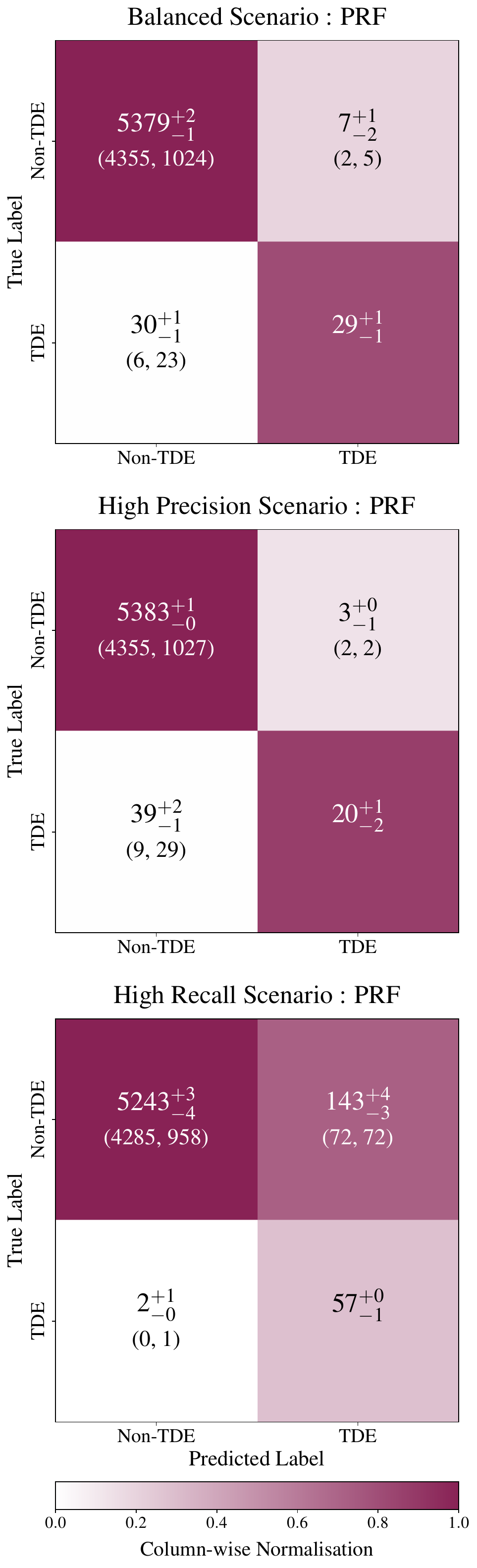}
        \caption{}
        \label{fig:PRF_CM}
    \end{subfigure}
    \hspace{1cm}
    \begin{subfigure}{0.36\textwidth}
        \centering
        \includegraphics[width=\textwidth,alt={Three XGBoost confusion matrices for the same three thresholds, recovering 28, 26 and 55 TDEs but with more false positives.}]{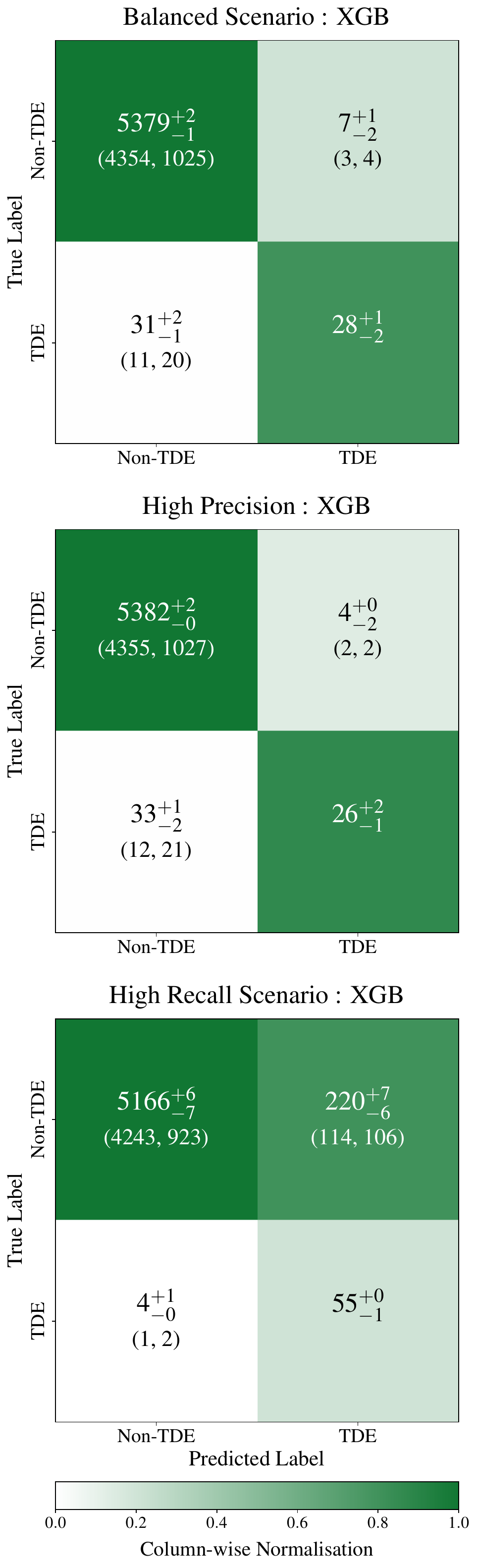}
        \caption{}
        \label{fig:XGB_CM}
    \end{subfigure}
    \caption{Confusion matrix plots for different threshold selection scenarios for both classifiers. For each blocks in the confusion matrix, the numbers in the bracket shows the counts of the AGN and SNe belonging to that particular block (There are 59 TDEs, 1029 SNE and 4357 AGN). The colors in the confusion matrix are normalized column-wise with the normalization scale shown at the bottom. Top panel shows the confusion matrices for the balanced scenario (minimum 80\% purity). The middle panel shows the scenario favoring high precision ($\approx$85\% purity). The bottom panel shows the scenario favoring the high recall scenario ($\approx$95\% completeness).}
    \label{fig:CMplots}
\end{figure*}

As briefly mentioned before, choice of a prediction threshold is required to convert the predicted probabilities of a classifier to binary labels for the metrics. Thus we perform threshold tuning to monitor the progression of these metrics to select the optimum threshold for the scenarios we are interested in. The results of the threshold tuning for both classifiers are shown in Figure \ref{fig:thresh}. Following the work of \cite{Stein24}, we define three scenario for threshold selection : Balanced scenario (minimum 80\% purity), High Precision scenario ($\approx$85\% purity), High Recall scenario ($\approx$95\% completeness, XGBoost does not reach this value hence the closest is reported). We select a cut of $\approx$ 85\% for the high precision case, in comparison to 95\% in \cite{Stein24}, since both of our classifier do not hit this mark in a reasonable threshold limit. The metrics for these scenarios for each classifier and the corresponding confusion matrices are shown in  Table \ref{tab:metrics} and Figure \ref{fig:CMplots} respectively.

Similar to the stability plots (Figure \ref{fig:violinplots}), the confusion matrices display the careful and conservative nature of PRF compared to XGBoost across all three scenarios. XGBoost outperforms PRF with $\approx$2\% higher recall in the balanced scenario, and $\approx$9\% higher recall in the high precision scenario. This performance gap stems from the binary nature of the XGBoost classifier, which tends to drive probability scores toward extremes ($P\approx0$ or $1$), effectively locking in candidates that match the training set features. While this allows XGBoost to maintain high completeness in high-SNR regimes by aggressively classifying TDE-like shapes, it lacks uncertainty awareness. By exhibiting this bimodal nature, XGBoost often assigns near-zero probability to ambiguous TDEs with AGN-like or SN-like traits, which effectively removes them from the sample regardless of threshold tuning. In contrast, PRF, by incorporating feature uncertainties, results in relatively broader probability distributions for lower quality data. Although this sensitivity to measurement noise causes valid but ambiguous TDEs to fall below strict high-precision cuts, it serves as a safety valve against artifacts and highlights its more conservative approach. This advantage is evident in the high-recall scenario, where both classifiers reach a completeness of $\approx$95\%. Here, PRF achieves significantly higher efficiency, rejecting $\approx$35\% more non-TDEs (143 FPs vs. 220 FPs) than XGBoost. These rejected sources are likely non-TDEs that the XGBoost model overfits as TDEs, but which PRF correctly penalizes due to the higher uncertainties in their features.

\subsection{Feature importance}\label{subsec:SHAP}

\begin{figure*}
    \centering
    \includegraphics[width=\linewidth, alt={Back-to-back SHAP importance bars for PRF and XGB, linked by ribbons; PRF ranks pre-peak variability first, XGB ranks peak temperature first.}]{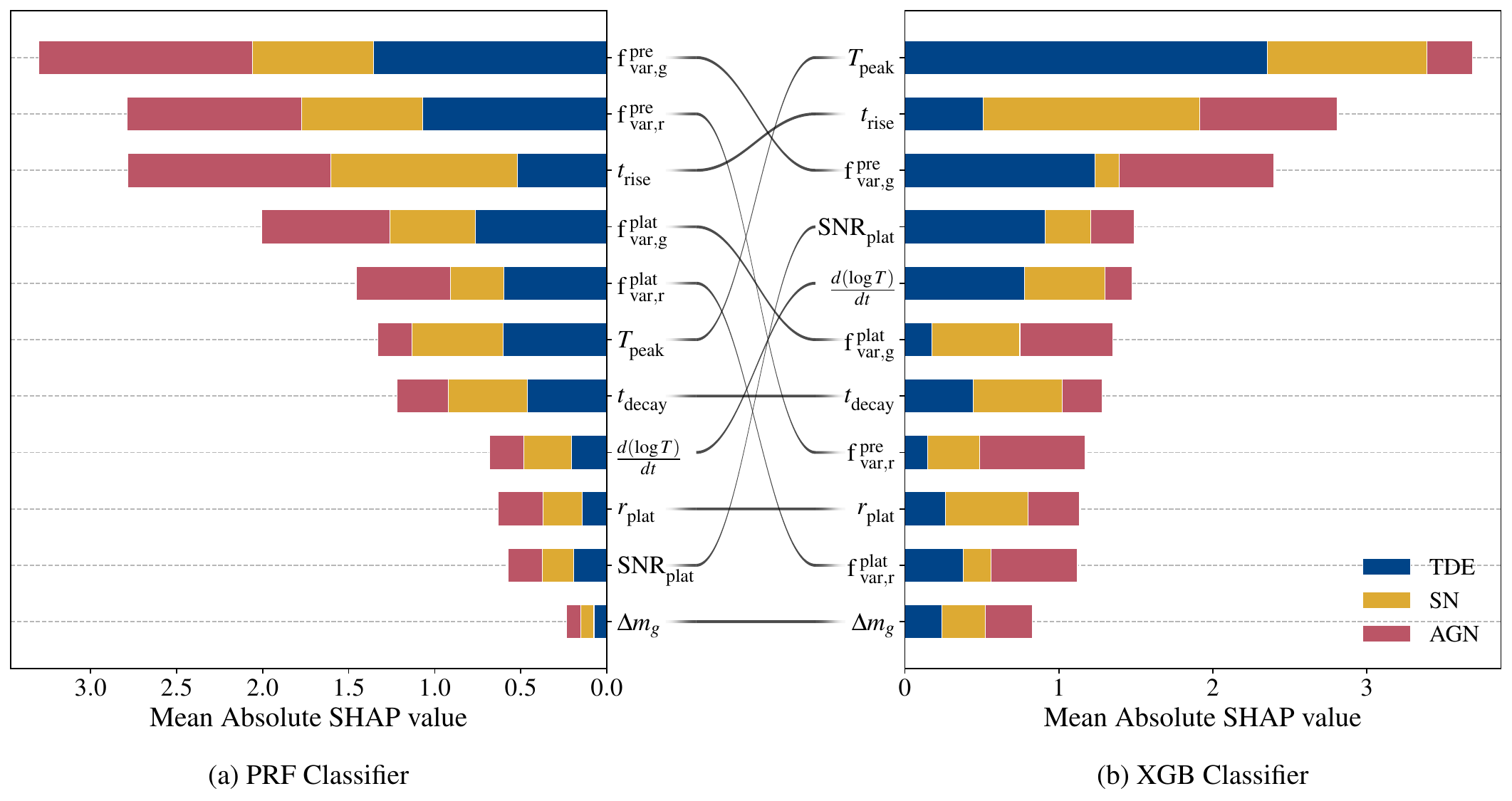}
    \caption{This figure shows the global feature importance for both classifiers. The features are ranked based on their global mean absolute SHAP values, which represent the average contribution of each feature to the model's predictions across the dataset. Each bar is divided into three colored segments corresponding to the three classes: TDE (blue), SN (orange), and AGN (red). The length of each segment indicates the average contribution of that feature to the prediction of the respective class. }
    \label{fig:feat}
\end{figure*}

To understand the decision processes of each classifier and to verify whether they are based on sound reasoning, we calculate both local and global feature importances using the \texttt{SHapley Additive exPlanations} (SHAP) python package \citep{lundberg_unified_2017}. The global feature importance plot for both classifiers are shown in Figure \ref{fig:feat}. 

For XGBoost, feature contributions are computed using \texttt{TreeExplainer}, which directly evaluates the internal structural paths of the gradient-boosted trees. The incorporation of uncertainties within the PRF architecture prevents the use of standard tree explainers, requiring feature contributions to be extracted using a custom Kernel SHAP estimator. Our custom brute force version evaluates the model over all possible feature subsets, replacing non-selected features with background values, and recovers per-feature contributions via a weighted linear fit. The feature values and their measurement uncertainties are perturbed jointly across all subsets, preserving the coupled structure of the model inputs. 

PRF identifies $\rm f_{\rm var,g}^{\rm pre}$ as its primary global feature, whereas XGBoost assigns it the third rank, with both models using the parameter to distinguish stochastic AGN variability from TDEs. Due to the independent parallel learning characteristic of bagging algorithms, the PRF retains the physically redundant $\rm f_{\rm var,r}^{\rm pre}$ as its second most important feature. In contrast, the sequential boosting architecture of XGBoost statistically penalizes this redundancy by demoting the feature to the eighth position. The plateau-phase fractional RMS features, $\rm f_{\rm var,g}^{\rm plat}$ and $\rm f_{\rm var,r}^{\rm plat}$, follow similar importance patterns across both models primarily using it to distinguish TDE and AGN. The prioritization of RMS features within the PRF relative to XGBoost stems from fundamental differences between bagging and boosting architectures. Because bagging relies on the ensemble averaging of independent decision trees, the architecture prioritizes global utility, which pushes the ranking of RMS features due to the high frequency of AGN within the training set. XGBoost minimizes residuals sequentially, allowing the initial splits to focus on global features while subsequent trees prioritize the local utility of features to reduce bias across specific subsets of the data.  

Both classifiers use the rise timescale ($t_{\rm rise}$) with high priority, with XGBoost ranking it second and PRF ranking it third, where it is used particularly to separate SNe from the other nuclear transients. The decay timescale ($t_{\rm decay}$) is ranked seventh by XGBoost and PRF, suggesting it as a weaker feature due its higher overlap among different classes relative to $t_{\rm rise}$. Moving on to the temperature parameters, XGBoost utilizes both of them significantly higher compared to PRF. XGBoost ranks $T_{\rm peak}$ globally first (and most important for TDEs) to isolate the hotter TDE population from SNe. In contrast to this, PRF ranks $T_{\rm peak}$ sixth, even though it learns a similar pattern. This can be explained by the fact that the multi-band modelling required to estimate $T_{\rm peak}$ introduces larger uncertainties that PRF accounts for but which XGBoost is blind to, combined with the tendency of the bagging  architecture discussed above to distribute discriminative weight more evenly across features. Together, these factors suppress the feature's global information gain in the PRF relative to other parameters, in contrast to XGBoost. Similarly, the temperature gradient ($d (\log_{10}T)/dt$) is ranked higher (fifth) by XGBoost using it to differentiate TDEs and SN which shows clear differences in temperature evolution. Meanwhile PRF ranks it eighth, where the same reasoning applies but with a weaker effect due to the higher variance of this parameter compared to $T_{\rm peak}$.

Turning finally to the plateau parameters, XGBoost ranks ${\rm SNR}_{\rm plat}$ fourth globally and third for TDEs, utilizing its boosting architecture to target the misclassified residuals of the specific TDE subpopulation containing late-time plateaus. Since the plateaus are present only for a fraction of TDEs, the PRF architecture (due to previously discussed methodological differences) fails to learn this sparse parameter as a strong indicator of TDEs, reducing its rank to tenth.

In summary, XGBoost operates as a relatively more physical optimizer, leveraging point estimates to construct rigid decision boundaries around physically motivated features. This maximizes its discriminative power for localized transient physics, though it operates entirely blindly to feature errors. PRF operates as a conservative probabilistic evaluator that is less reliant on localized physics. It prioritizes stable and low variance metrics that generalize across the population, avoiding over-reliance on features affected by large feature uncertainties. The implications of this architectural difference for the low-SNR regime expected from the  Rubin Observatory are discussed further in Section \ref{subsec:SNRcomp}.

\section{Applying classifiers on real data}
\subsection{Testing on real TDEs}\label{subsec:realTDEtest}

\begin{table*}
    \renewcommand{\arraystretch}{1.3}
    \centering
    \begin{threeparttable}
    \begin{tabular}{p{16em}|cc|cc}
        \hline \hline
        \multirow{2}{*}{Name} & \multicolumn{2}{c|}{PRF} & \multicolumn{2}{c}{XGB} \\
        \cline{2-5}
        & $P(\rm{TDE})$ & Next-highest class & $P(\rm{TDE})$ & Next-highest class \\
        \hline \hline
        ZTF24aajvvhj / TDE2024gre\tnote{a} & $0.975\pm0.002$ & SN (0.020) & $1.000\pm0.000$ & -- \\
        
        ZTF23abvzeqp / TDE2024kt\tnote{b} & $0.974\pm0.003$ & SN (0.020) & $1.000\pm0.001$ & -- \\

        ZTF23abkixdb / TDE2023wdb\tnote{c} & $0.967\pm0.003$ & SN (0.025) & $1.000\pm0.000$ & -- \\

        ZTF23abohtqf / TDE2023xen\tnote{d} & $0.966\pm0.005$ & SN (0.026) & $0.996\pm0.010$ & AGN (0.003) \\

        ZTF23aapyidj / TDE2023mfm\tnote{e} & $0.959\pm0.009$ & SN (0.033) & $1.000\pm0.000$ & -- \\

        ZTF24aaecooj / TDE2024bgz\tnote{f} & $0.945\pm0.013$ & SN (0.047) & $1.000\pm0.000$ & -- \\

        ZTF23abgnxfv / TDE2023tmq\tnote{g} & $0.933\pm0.011$ & SN (0.058) & $0.988\pm0.012$ & AGN (0.008) \\

        ZTF20abwtifz / TDE2020afhd\tnote{h} & $0.883\pm0.027$ & SN (0.101) & $0.982\pm0.058$ & SN (0.017) \\

        ZTF23abaujuy / TDE2023rvb\tnote{i} & $0.868\pm0.014$ & SN (0.077) & $0.957\pm0.134$ & AGN (0.033) \\

        ZTF18aabuoxd / TDE2018mli\tnote{j} & $0.854\pm0.020$ & SN (0.107) & $1.000\pm0.000$ & -- \\

        ZTF24aakaiha / TDE2024gxr\tnote{k} & $0.692\pm0.028$ & AGN (0.271) & $0.085\pm0.242$ & AGN (0.914) \\

        ZTF23aaqdjhi / TDE2023mhs\tnote{l} & $0.007\pm0.002$ & SN (0.979) & $0.000\pm0.000$ & SN (1.000) \\
        \hline
        \hline
    \end{tabular}
    \begin{tablenotes}[para,flushleft]
    \fontsize{8}{8}\selectfont
        TNS classification report/publications: \item[a] \cite{Somalwar2024} 
        \item[b] \cite{Poidevin2024}
        \item[c] \cite{Patra2023} 
        \item[d] \cite{Somalwar2023b} 
        \item[e] \cite{Chornock2023, Li2026} 
        \item[f] \cite{Godson2024}
        \item[g] \cite{Somalwar2023a}
        \item[h] \cite{Hammerstein2024} 
        \item[i] \cite{Hammerstein2023}
        \item[j] \cite{Arcavi2023}
        \item[k] \cite{Fremling2024}
        \item[l] \cite{Sollerman2023}
    \end{tablenotes}
    \caption{Photometric classification of known TDEs, detected after mid-2023, which therefore have not been included in the training data.  We show the predicted TDE probability and the dominant non-TDE class (SN or AGN, with its probability in parentheses; the 68\% spread is computed using randomized SMOTE oversampling of the training data). Rows are ordered by the PRF TDE probability. A dash indicates that neither non-TDE class exceeds 0.001. The reported probabilities are uncalibrated classifier outputs.}
    \label{tab:realTDE_test}
\end{threeparttable}
\end{table*}

Since the nuclear transient filter (Section \ref{subsec:nuc_dataset}) was operated until mid-2023, the training set is restricted to the labels obtained prior to this cutoff. This constraint allows us an extra validation step, enabling us to evaluate the classifiers on TDEs discovered after the cutoff. We queried TNS for spectroscopically classified TDEs discovered between mid-2023 and mid-2024, allowing a 1.5-year window for estimating the plateau features from the lightcurves. We perform the same cuts as for the training data, resulting in a total of 12 sources. To classify these recent TDEs, we train an ensemble of models on the complete dataset  \red{(since we are testing new data, we can use the entire dataset without the stratified fold + splitting used in Section \ref{subsubsec:LOOCV}, which then contains 59 TDEs, 1029 SNe and 4357 AGN)} using different random seeds for the SMOTE algorithm, allowing us to derive classification uncertainties that reflect the internal randomness and stability of the pipeline. The TDEs used for testing and the corresponding predicted probabilities are shown in Table \ref{tab:realTDE_test}.

For both classifiers, 10 out of the 12 TDEs have a predicted probability greater than 0.85 (the high precision scenario), indicating that the extracted features effectively discriminate TDEs from other transient types. This recovery rate is consistent with the recall of each models at these predicted probabilities (0.56 $\pm$ 0.02 for PRF; 0.58 $\pm$ 0.03 for XGBoost), which predicts $\sim$ 7–8 recovered TDEs out of 12. Recovering ten or more is compatible with this expectation (one-sided binomial test, p value $\approx$ 0.06). The modest excess over the expected recovery may suggest that these TDEs are intrinsically easier to classify, though with twelve events it is equally consistent with a statistical fluctuation. The probabilities and their associated error bars show how the training data affects the stability of each classifier, with XGBoost producing highly confident predictions for the majority of sources while PRF provides high probabilities that include a broader margin for doubt.

The two TDEs predicted with much lower probabilities are TDE2024gxr (PRF $0.69\pm0.03$; XGBoost $0.09\pm0.24$) and TDE2023mhs (PRF $0.007\pm0.002$; XGBoost $0.00\pm0.00$). Using local feature importance plots (generated using SHAP (section \ref{subsec:SHAP})), XGBoost strongly classifies TDE2024gxr as an AGN, primarily driven by the pre-transient fractional rms parameters. Using the higher uncertainties in these parameters, PRF assigns a much higher probability resulting in a better classification compared to XGBoost. Both classifiers strongly identified TDE2023mhs as a supernova, with SHAP plots indicating   $T_{\rm{peak}}$, $t_{\rm{decay}}$, $t_{\rm rise}$ and $d(\log{T})/dt$, being the primary parameters driving this decision. This source has a rapid rise and decay timescale (one of the fastest observed TDEs), which creates ambiguity for the models, suggesting an SN-like classification. While the absence of $r$-band data at the peak for this source led to miscalculated lower temperatures, sensitivity tests by manually inflating the uncertainties of the temperature parameters only increases the TDE probability to a maximum of 12\%. This indicates that while the poorly constrained temperature contributes to the misclassification, the primary driver remains the structural similarity of the light-curve timescales to typical supernova profiles.

We note that the probabilities quoted throughout this section are uncalibrated classifier outputs. Their relative ordering is meaningful, but their absolute values should not be interpreted as posterior probabilities of the true class. A formal calibration analysis is beyond the scope of this work.

\subsection{Searching for new TDEs}\label{subsec:newTDEsearch}

\begin{table*}
    \renewcommand{\arraystretch}{1.25}
    \centering
    \begin{threeparttable}

    \begin{tabular}{m{12em}|m{7em}|m{7em}|m{6em}|m{5.5em}|m{7em}}
    \hline \hline
       Name & Redshift & IR Echo & Host W1-W2 & Late-time Plateau & Peak Magnitude\\
        \hline 
        \multicolumn{6}{c}{\textbf{Candidates from Unknown sources}}\\
        \hline
        \vspace{0.2em}
        ZTF22aajmjxs / AT2022jpt\tnote{a} & $0.11\pm0.09$ $\rm (P^{\square})$ & Insufficient data & - & Yes & 19.52\\
        ZTF22aaimgkp / AT2022jks\tnote{b*}& $0.33\pm0.04$ $\rm (P^{\triangle})$ & Insufficient data & 0.641 $\pm$ 0.142 & Still decaying & 19.13\\
        ZTF22abvbjkl / AT2022aaun\tnote{c}& $0.20\pm0.03$ $\rm (P^{\triangle})$ & No & 0.381 $\pm$ 0.15 & No & 19.56\\
        ZTF21aaxtqvj / AT2021lwh\tnote{d*}& $0.10\pm0.03$ $\rm (P^{\triangle})$ & Insufficient data & - & No & 20.09\\
        ZTF21aaqardq / AT2021gkm\tnote{e}& $0.14\pm0.02$ $\rm (P^{\triangle})$ & No & 0.282 $\pm$ 0.12 & No & 19.81\\
        ZTF21aacsvko / AT2021asq\tnote{f}& $0.20\pm0.03$ $\rm (P^{\triangle})$ & No & 0.398 $\pm$ 0.124 & Yes & 19.67\\
        ZTF21acojhgu / AT2021aees\tnote{g*}& $0.29\pm0.01$ $\rm (P^{\triangle})$ & Yes & 0.257 $\pm$ 0.125 & Yes & 19.02\\
        ZTF20acpmojv / AT2020zev\tnote{h}& $0.13\pm0.02$ $\rm (P^{\triangle})$ & No & 0.170 $\pm$ 0.171 & No & 20.10\\
        ZTF18aczegek\tnote{**} & $0.29\pm0.05$ $\rm (P^{\square})$ & Insufficient data & 0.036 $\pm$ 0.108 & Yes & 19.26\\
        ZTF19acanuza & - & No data & 0.020 $\pm$ 0.146 & No & 20.14\\
        ZTF20aatpzog\tnote{*} & $0.62\pm0.04$ $\rm (P^{\triangle})$ & Yes & 0.287 $\pm$ 0.099 & No & 19.84\\
        \hline

        \multicolumn{6}{c}{\textbf{Candidates from Spectroscopic SN sources}}\\
        \hline
        \vspace{0.2em}
        ZTF19abclykm / SN2019meh (SLSN-II)\tnote{i} & 0.094 (S) & Yes & 0.656 $\pm$ 0.031 & No & 17.51\\
        \hline
        \multicolumn{6}{c}{\textbf{Candidates from classified AGN sources}}\\
        \hline
        \vspace{0.2em}
        ZTF20acxtaau / AT2020actc\tnote{j}& $0.1454$ (S) & Maybe & 0.605 $\pm$ 0.047 & Yes & 19.09\\
        ZTF19aavwtcb / AT2019gtm\tnote{k}& 0.0485 (S) & Yes & 0.132 $\pm$ 0.085 & Yes & 18.80\\
        \hline \hline

    \end{tabular}
    \begin{tablenotes}[para,flushleft]
    \fontsize{8}{8}\selectfont
        TNS discovery/classification report: \item[a]  \cite{Munoz-Arancibia2022}
        \item[b] \cite{Fremling2022}
        \item[c] \cite{Forster2022}
        \item[d] \cite{Gompertz2021}
        \item[e] \cite{2021TNSTR.834....1M}
        \item[f] \cite{Tonry2021}
        \item[g] \cite{Munoz-Arancibia2021}
        \item[h] \cite{2020TNSTR3415....1F}
        \item[i] \cite{Nicholl2019}
        \item[j] \cite{Munoz-Arancibia2020}
        \item[k] \cite{Nordin2019}
    \end{tablenotes}
    \begin{tablenotes}[para,flushleft]
    \fontsize{8}{8}\selectfont
        \item[*] detected in Swift/UVOT
        \item[**] detected in Swift/UVOT during plateau phase, but no significant detection above the host level
    \end{tablenotes}
    \caption{New TDE candidates identified by the search conducted on all nuclear transients spanning from 2018 to mid-2023, grouped by their original classification: unknown nuclear transients,  spectroscopically classified supernovae, and known AGN. We list the ZTF identifier and the IAU name (if available on TNS). For supernovae we report the spectroscopic classification, and corresponding TNS report, in brackets. The \textit{Redshift} column reports photometric redshifts derived from the Legacy Survey \citep{Zhou2025} ($\rm P^{\triangle}$) or, where unavailable, from SDSS \citep{Bolton2012} ($\rm P^{\square}$), and spectroscopic redshifts where available (S). The \textit{IR Echo} column indicates whether a mid-infrared echo was detected in the WISE light curves \citep{Wright2010}. The \textit{Host W1-W2} column reports the WISE mid-infrared color of the host galaxy and the \textit{Late-time Plateau} column indicates whether a persistent late-time plateau is present in the optical light curve. The \textit{IR Echo} and \textit{Late-time Plateau} assessments were performed through manual inspection of individual sources. The \textit{Peak Magnitude} column reports the measured AB magnitude of the peak of the transient in the $r$ band.}
    \label{tab:newTDEsearch}
    \end{threeparttable}
\end{table*}

\begin{figure*}
    \centering
    \begin{subfigure}{0.49\textwidth}
        \centering  
        \includegraphics[width=\textwidth, alt={Corner plot of log rise time, peak temperature and decay time showing new candidates overlapping the known TDEs.}]{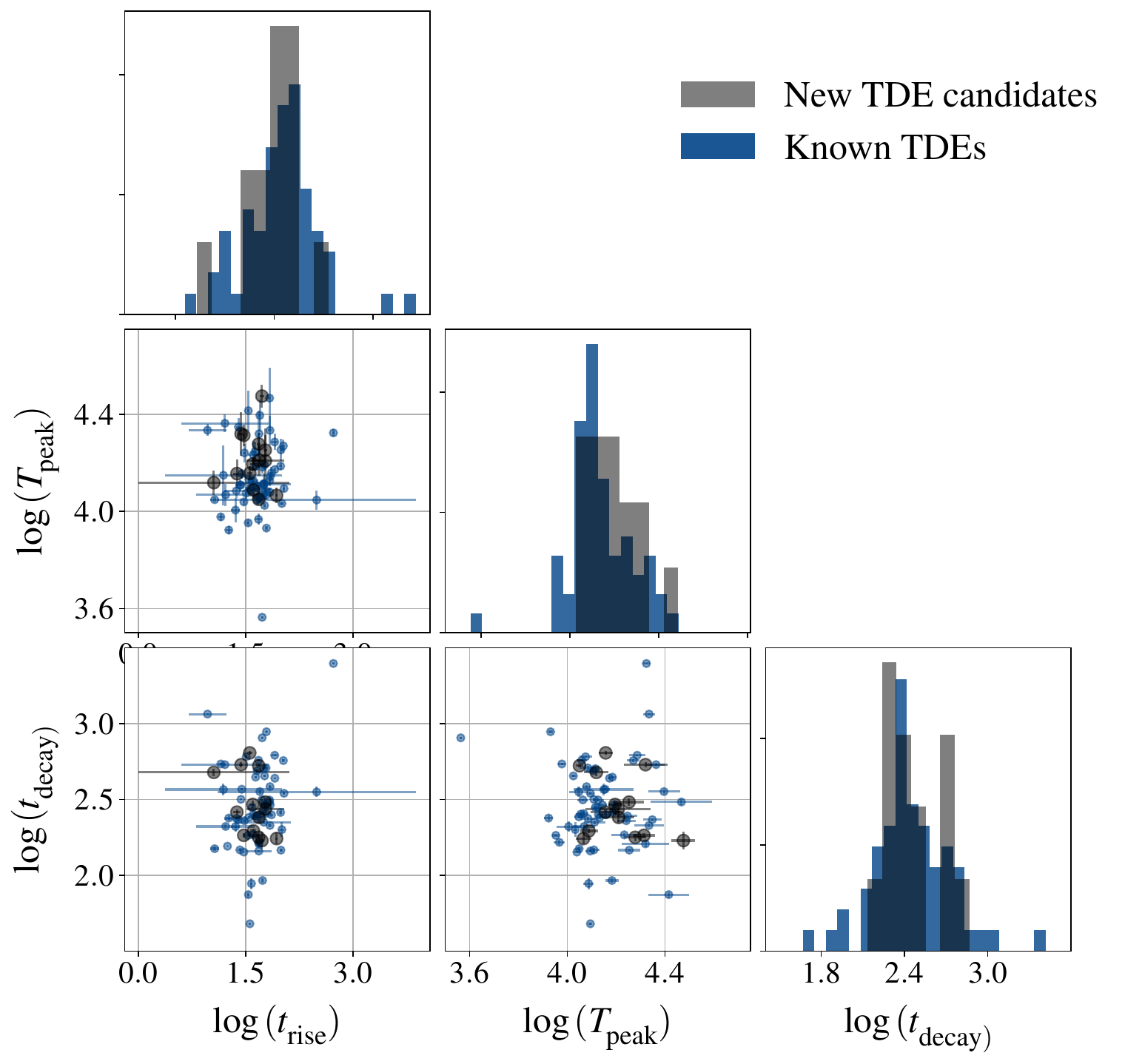}
        \caption{}
        \label{fig:cornerUNK}
    \end{subfigure}
    \begin{subfigure}{0.46\textwidth}
        \centering  
        \includegraphics[width=\textwidth, alt={Stacked histogram of peak AB magnitude; new candidates fall at the faint end of the known-TDE distribution.}]{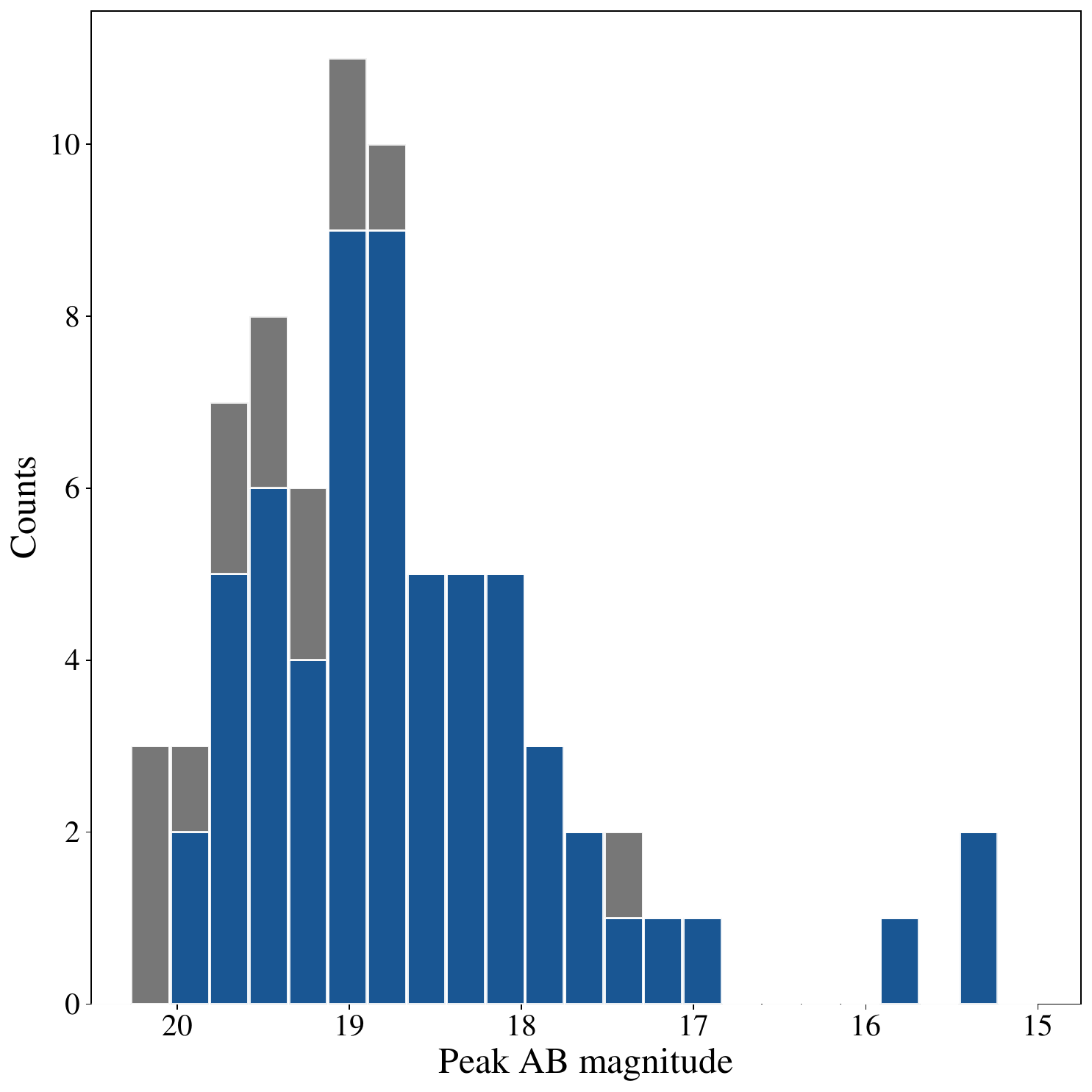}
        \caption{}
        \label{fig:peakmag_unk}
    \end{subfigure}
    \caption{This figure compares the properties of the known TDE training sample to those of the newly identified candidates. The left panel displays the inferred physical parameter distributions ($t_{\rm rise}$, $T_{\rm peak}$, $t_{\rm decay}$), demonstrating the morphological overlap between the two populations. The right panel presents the stacked distribution of peak AB magnitudes for both the known TDEs and the new candidates. The new candidates cluster at fainter magnitudes, indicating that the selection algorithm does not favor brighter transients alone and successfully identifies sources outside the known brightness distribution. \red{Note that the distributions in the left panel are overlaid to show the overlap between the two populations, while those in the right panel are stacked.}}
    \label{fig:unk_plots}
\end{figure*}

Following the evaluation of the classifier performances, as a proof-of-concept, we apply the models to search for new candidate TDEs within the nuclear transient data. The PRF and XGBoost classifiers are architecturally complementary: PRF penalizes sources with high feature uncertainties while XGBoost enforces rigid physically motivated boundaries. Requiring both classifiers to independently label a source as a TDE constructs a conservative orthogonal filter that guards against the distinct failure modes of each architecture. Equal weighting is adopted as a default conservative choice since determining the optimal weighting strategy based on classifier calibration is left for future work. 

We conduct this search in two parts. The first part searches for new candidates among the unknown nuclear transients by training both models on the \red{entire} dataset and applying the predetermined high precision threshold.  The second part re-evaluates sources in the existing training data to locate early-time misclassifications and instances of overlapping classes, such as TDEs present in an AGN. Because these classified SNe and AGN are already part of the training data, applying the full-dataset model directly would introduce training bias and invalidate the evaluation. To prevent this, the second approach employs the leave-one-out cross-validation methodology, where a model iteration is trained by excluding the specific source under evaluation. Paired with the high precision threshold, this method captures sources that share properties with non-TDE classifications. We found 11 candidates from the unknown sources, 1 candidate from classified SNe, and 2 from known AGN. The results of this search are presented in Table \ref{tab:newTDEsearch}. \red{In addition to the candidates selected by both the classifiers, there are 7 more candidates\footnote{ZTF18acbwomm, ZTF20abbpxut, ZTF20abkstjk, ZTF20achupkw (SLSN-II), ZTF21aabgjcz, ZTF21aaxtlvc, ZTF21aazenvp} (6 unknowns and 1 SN) only selected by the PRF classifier and 12 more candidates \footnote{ZTF18aadbtae, ZTF18acvwkvc, ZTF19aaciohh (AGN), ZTF19abiaxpi, ZTF19abuonfl, ZTF19aclocml, ZTF19acssksf, ZTF19acuywme, ZTF20acpgnmw, ZTF21abavbci, ZTF22aajnyhg, ZTF22abwvamn} (11 unknowns and 1 AGN) only by the XGB model.  The per-source probabilities from both classifiers are available in the data release (Section \ref{sec:dataavail}), allowing these selections to be reproduced at any threshold. In this work, we stay with the ones classified by both independently.  }

At the high precision filter , the ensemble reaches a precision of 0.90 and a recall of 0.32, with a false positive rate of $3.7\times10^{-4}$. This represents a modest precision gain and a lower false positive rate relative to the individual classifiers, at the expected cost of recall. We do not calculate a purity for the 11 candidates recovered from the unknown sources, as the precision is estimated on the training data and need not carry over to a population with a different underlying class composition. The false positive rate is instead a property of the classifier, and predicts $\approx2$ false positives from the 6030 unknown nuclear transients that were screened, suggesting that the 11 candidates are unlikely to be dominated by false positives. 

For the candidates recovered from the re-evaluation of the labelled data, the ensemble predicts false positives of the order of 0.4 SNe and 1.6 AGN across the 1029 SNe and 4357 AGN evaluated. The one SN and two AGN recovered are consistent with these expected counts and do not by themselves indicate a misclassification. Individual inspection of these sources, however, support the hypothesis that these are TDEs misclassified in the labeled data. 
\begin{itemize}
    \item[$\bullet$] ZTF19abclykm / SN2019meh is spectroscopically classified as a SLSN-II \citep{Nicholl2019}. However the second peak in its light curve (Fig. \ref{figA:LCplots}) challenges this classification. Similar SLSN misclassifications have been reported by \citet{Frederick2021}.  
    
    \item[$\bullet$] ZTF20acxtaau / AT2020actcj is classified as an broad-line AGN based on its SDSS spectrum \citep{Liu19, Flesch2023}. The host galaxy is detected at radio wavelengths (1.7 mJy in FIRST \citep{Becker1995}). The post-peak ZTF data shows a significant plateau with no evidence for additional variability, making this a strong candidate TDE in an AGN.  

\item[$\bullet$] ZTF19aavwtcb / AT2019gtm
    is classified as an narrow-line AGN based on its SDSS spectrum \citep{Flesch2023}. The pre-peak WISE colors are consistent with a quiescent galaxy, while a dust echo is detected after the optical peak. Similar to AT2020actc, the post-peak plateau in ZTF  supports a TDE origin for this flare.

\end{itemize}

Figure \ref{fig:cornerUNK} compares the inferred physical parameter distributions ($t_{\rm rise}$, $t_{\rm decay}$, $T_{\rm peak}$) of the newly identified candidates against the confirmed TDE training sample. The candidate parameters exhibit high overlap with the established TDE distribution, demonstrating that the classifier isolates sources morphologically similar to the existing sample. Figure \ref{fig:peakmag_unk} details the peak AB magnitude distribution of the confirmed TDEs versus the   new candidates. The new candidates cluster at fainter magnitudes, confirming that the  host-agnostic, brightness-independent framework successfully recovers TDE-like sources outside the brightness range of the training population. The candidates identified in this search represent promising targets for follow-up observations to confirm their nature and further expand the sample of known TDEs.

\subsubsection{Comparison with similar searches}
\label{subsubsec:archival_comparison}

Several archival searches for TDEs in the ZTF stream have produced candidate lists that can be compared against our own predictions at the level of individual sources.

\citet{quintinLostFoundGallery2025} recovered a list of 23 nuclear transients from the ZTF archive using the Fink broker early TDE detection module \citep{lanzaEarlyIdentificationOptical2026}. The 23 candidates are further divided into 4 classes :  TDE, ANT, ambiguous sources and supernovae. Ten of their 19 TDE, ANT, and ambiguous sources fall outside the temporal reach of our search. Among the nine that remain, our ensemble classifier recovers ZTF20aatpzog through the unknown transient search and ZTF20acxtaau/AT2020actc through the re-evaluation of labeled AGN, both clearing our high-precision cut with a TDE preference from each classifier. five out of the remaining seven lie below that cut but are still ranked as TDEs by both of our classifiers over any other class and so remain viable candidates, and only two of them (ZTF20accxwrk and ZTF18aasvknh) are instead nudged towards AGN, a reading our classifiers offer in contrast with their TDE selection. Finally, only one of their four supernova candidates, AT2019agc/ZTF19aafmytc, appears in our dataset, and both of our classifiers independently read it as an SN.

\citet{pavez-herrera_alerce_2025} add a dedicated TDE subclass to the ALeRCE light curve classifier and publish a list of 56 archival candidates with a TDE probability above 50\%. From the list, when filtered for the temporal window and removing the already classified sources in our list, only 38 of them are in our unknown transient list. Of these, ZTF21aaqardq/AT2021gkm clears our high-precision cut with a TDE preference from both classifiers. Out of the remaining, 13 of them stay below the cut but maintain a TDEs preference from both classifiers, 17 of them are preferred as SN, and the remaining are discordant (one classifier leans TDE, the other SN or AGN). 

\citet{gomez_identifying_2023}  run the FLEET classifier over the ZTF archive and list 39 unclassified transients with a probability greater than 50\% ($P(\rm TDE) > 0.5$), where 29 of them are within the temporal window of our search. Two of them clear our high-precision cut with a TDE preference from both classifiers, ZTF19aavwtcb/AT2019gtm and ZTF21acojhgu/AT2021aees, the first among the re-evaluated labeled AGN and the later among the unknown nuclear transients. Nine more candidates from FLEET remain below the cut yet continue to be favored as TDEs by both classifiers, while the rest of the sources get classified as one of the other two classes (seven preferred as SN and five as AGN by both classifiers, while the remaining six are discordant).

While the candidate-level agreement across these independent searches is reassuring, a one-to-one comparison of the searches themselves remains hard, as each is built on a classifier operating at a different purity and completeness setting and uses different input information. Establishing such a comparison would require more curated common benchmarks, such as the MALLORN dataset \citep{magillMALLORNManyArtificial2025}, which provides a large set of simulated LSST light curves built from real ZTF nuclear transients, on which such searches can be evaluated on an equal footing.

\section{Discussion}\label{sec:discussion}

\subsection{Probabilistic vs. Deterministic Classification}\label{subsec:PRFvsXGB}

\begin{figure}
    \centering
    \includegraphics[width=\linewidth, alt={Ridgeline plot per TDE comparing probability densities; PRF distributions are narrow and unimodal, XGBoost ones broad and multi-peaked.}]{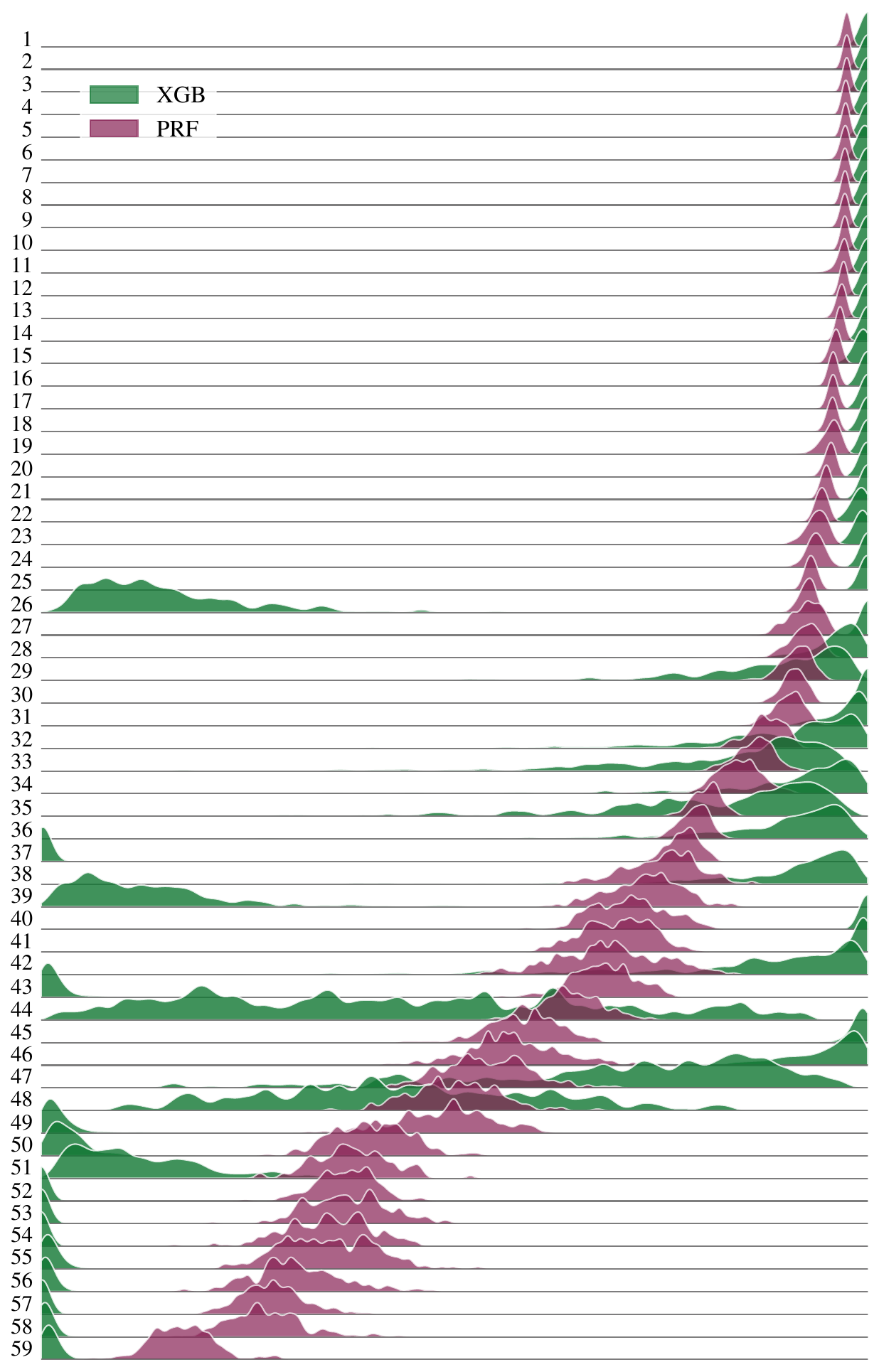}
    \caption{A detailed view of the TDE probability distributions originally presented in Figure \ref{fig:violinplots} across the 250 LOOCV repetitions. The distributions contrast the predictive stability of the models, detailing the unimodal probability spreads of the PRF classifier alongside the high-variance scattering of the XGBoost classifier for ambiguous sources.}
    \label{fig:PRF_XGB_comp_violin}
\end{figure}

Throughout the previous sections, the performance differences between the PRF and XGBoost classifiers highlight the critical impact of integrating measurement uncertainties, underscoring the difference between probabilistic and deterministic architectures. Figure \ref{fig:PRF_XGB_comp_violin} provides a detailed view of the TDE probability distributions across the 250 LOOCV repetitions. PRF maintains a stable, predominantly unimodal probability distribution that broadens marginally for lower-confidence sources, whereas XGBoost predictions for these same ambiguous sources scatter erratically across the entire probability domain. It is important to note that PRF's stability should not be confused with probabilistic calibration. Assessing whether the predicted probabilities accurately reflect true classification frequencies would require other methods, which is beyond the scope of this work.

This stability in PRF stems from two complementary architectural properties. First, the  bagging framework constructs independent trees in parallel and averages their predictions, reducing overall variance in the ensemble output and naturally reducing the variance introduced by LOOCV. Second, the native uncertainty propagation fundamentally  changes how each feature contributes to the classification. Rather than passing a single point estimate through each tree split, PRF integrates over the full posterior distribution of each feature measurement. For ambiguous sources with large measurement uncertainties, this integration distributes the classification weight across a broader range of parameter space, preventing any single imprecise measurement from dominating the prediction. The result is a conservative, broader probability distribution that reflects the data quality rather than forcing a confident decision. In the case of XGBoost, the extreme variance could be primarily linked to its structural mechanics. As a sequential gradient boosting algorithm, the removal or inclusion of a single ambiguous sample in a highly imbalanced dataset drastically shifts the loss gradient and the resulting global decision boundary. Adding to this, since XGBoost evaluates inputs as rigid point estimates  rather than probability density functions, it constructs sharp, localized decision boundaries that fail to generalize when LOOCV perturbations isolate boundary-case samples.

\begin{figure*}
      \centering
      \begin{minipage}[t]{0.49\textwidth}
          \centering
          \begin{subfigure}{\textwidth}
              \centering
              \includegraphics[width=\textwidth, alt={Light curve of TDE2022emf with median fit and posterior draws that fan out widely around the
  sparsely sampled rise.}]{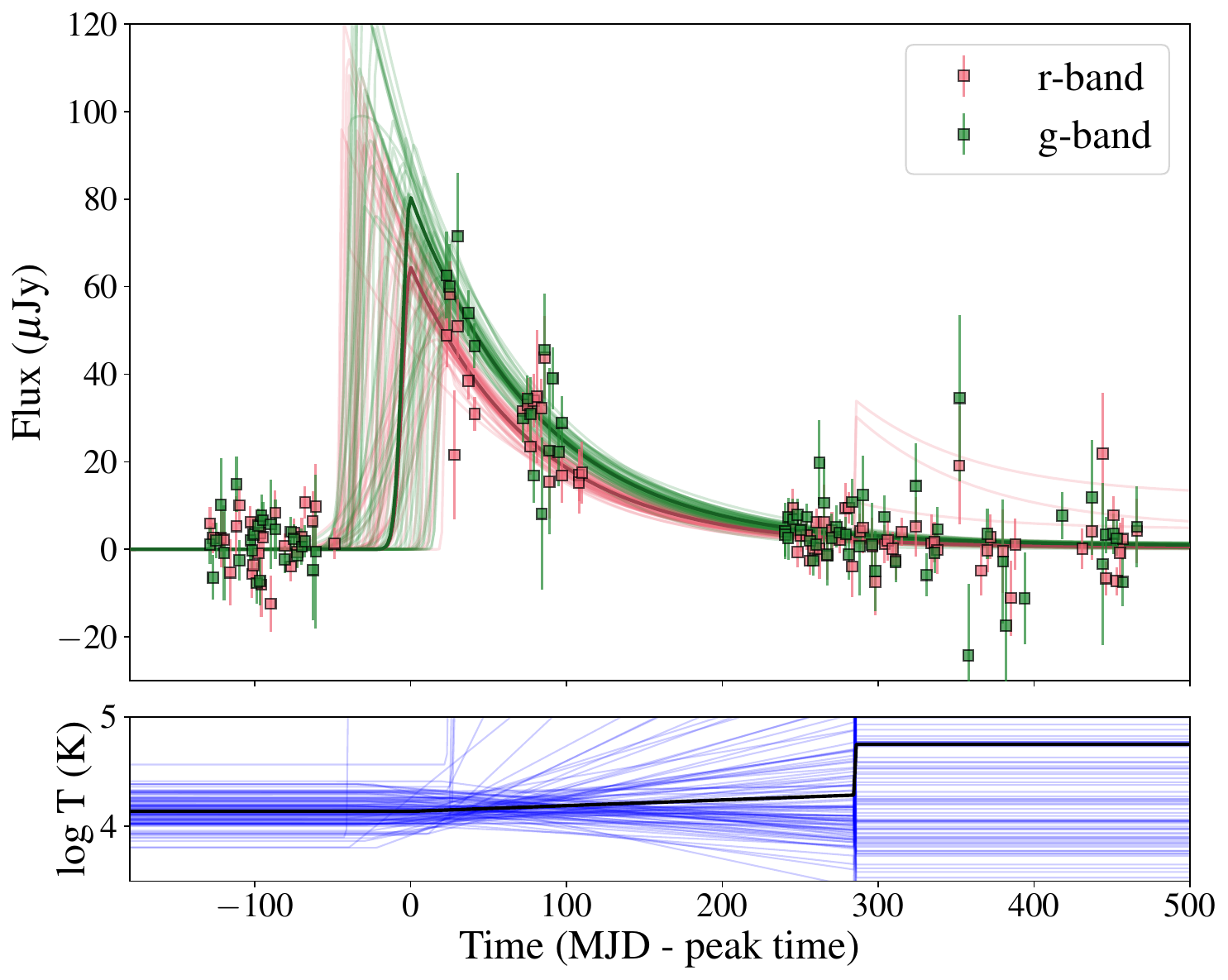}
              \caption{}
              \label{fig:PRFXGB_lcplot_TDE2022emf}
          \end{subfigure}\\[4pt]
          \begin{subfigure}{\textwidth}
              \centering
              \includegraphics[width=\textwidth, alt={Two SHAP waterfall plots for TDE2022emf, where XGBoost is driven by peak temperature with large
  opposing terms and PRF is driven by pre-peak variability with all contributions positive.}]{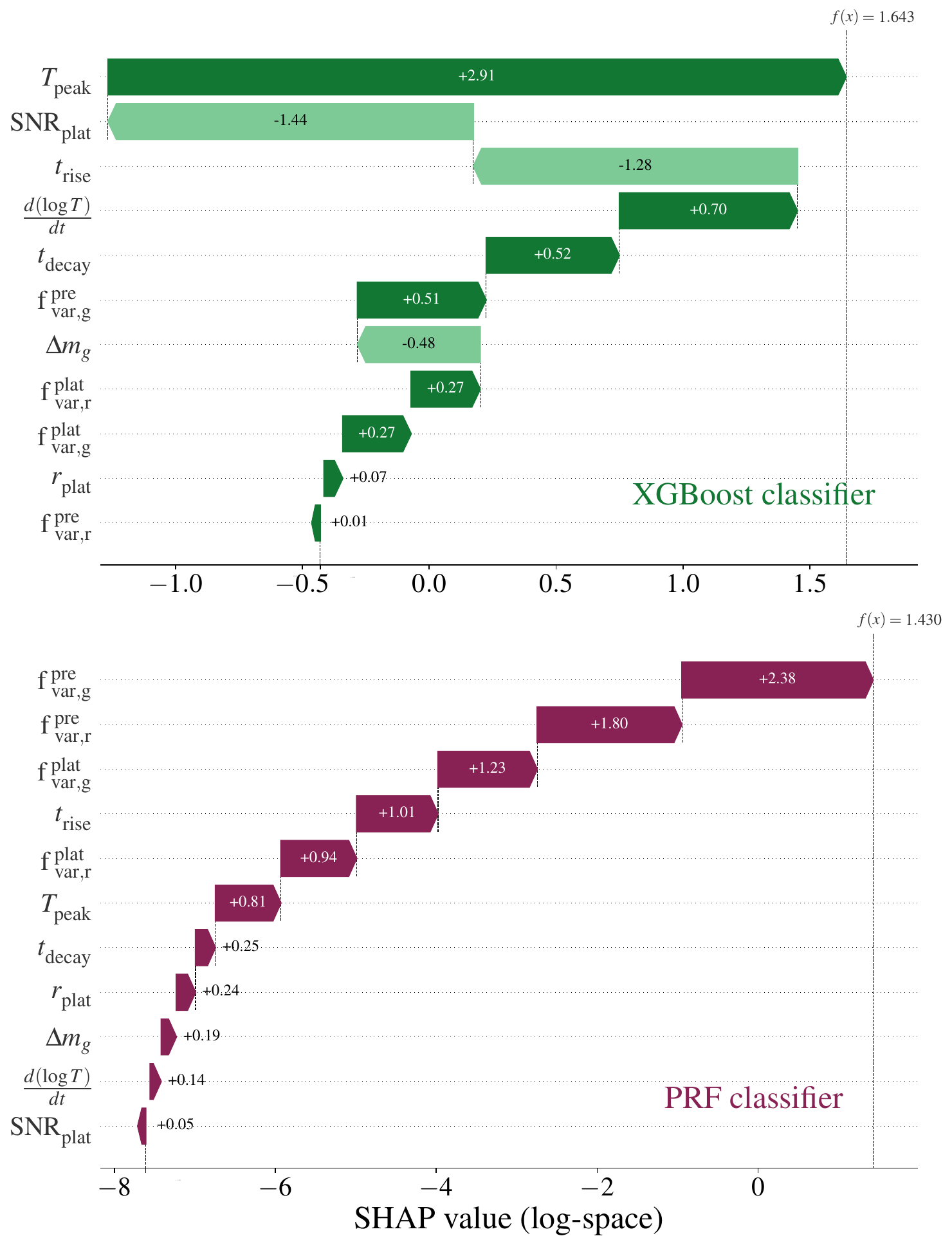}
              \caption{}
              \label{fig:prf_xgb_shap_comp_TDE2022emf}
          \end{subfigure}
      \end{minipage}\hfill
      \begin{minipage}[t]{0.49\textwidth}
          \centering
          \begin{subfigure}{\textwidth}
              \centering
              \includegraphics[width=\textwidth, alt={Light curve of TDE2022hvp, well sampled around the peak, with a lower panel showing the fitted
  temperature rising steeply to the 10^6 K limit before dropping at the plateau.}]{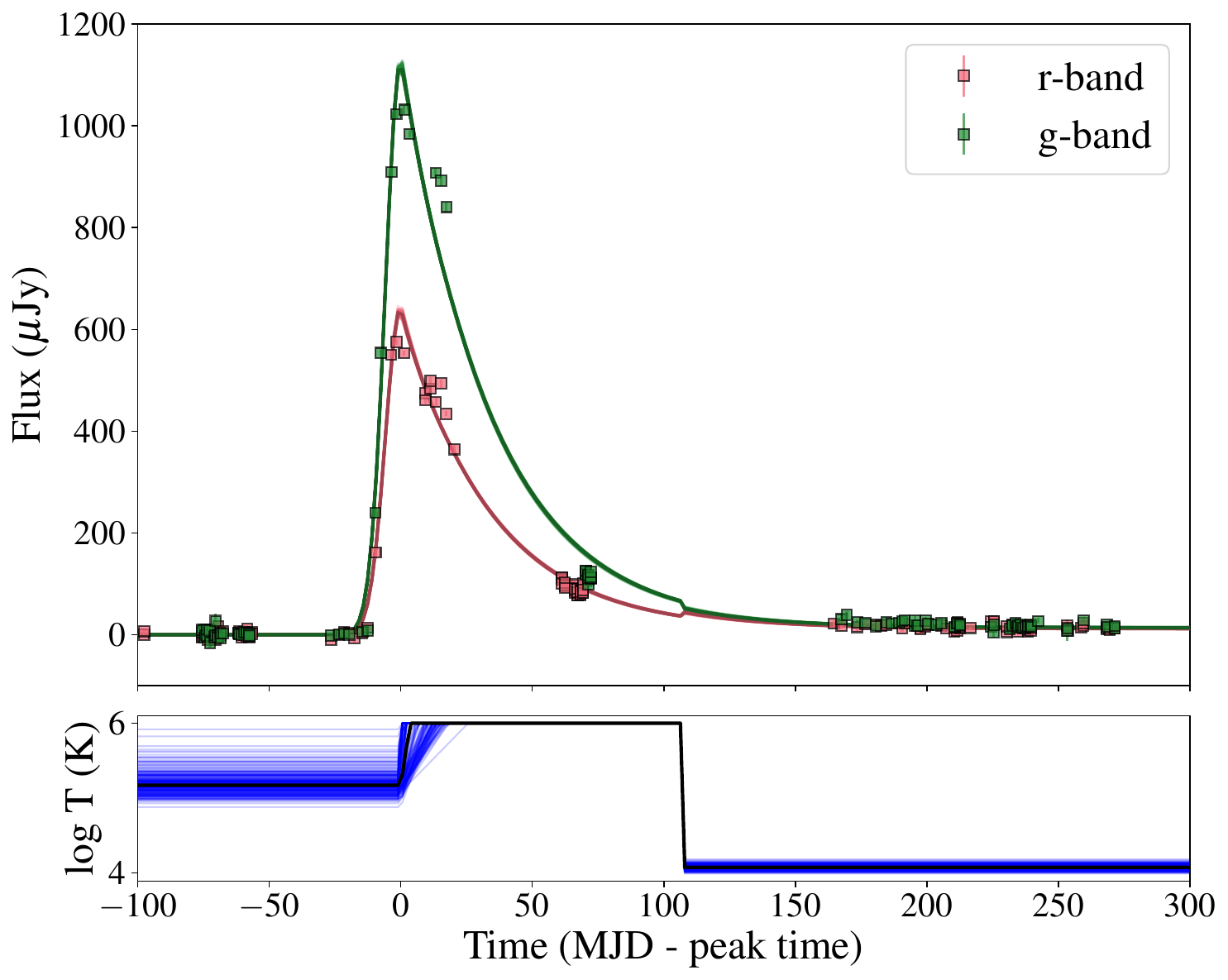}
              \caption{}
              \label{fig:PRFXGB_lcplot_TDE2022hvp}
          \end{subfigure}\\[4pt]
          \begin{subfigure}{\textwidth}
              \centering
              \includegraphics[width=\textwidth, alt={Two SHAP waterfall plots for TDE2022hvp, where XGBoost is dominated by a large negative
  temperature gradient term and PRF is driven by pre-peak variability with the temperature gradient ranked last.}]{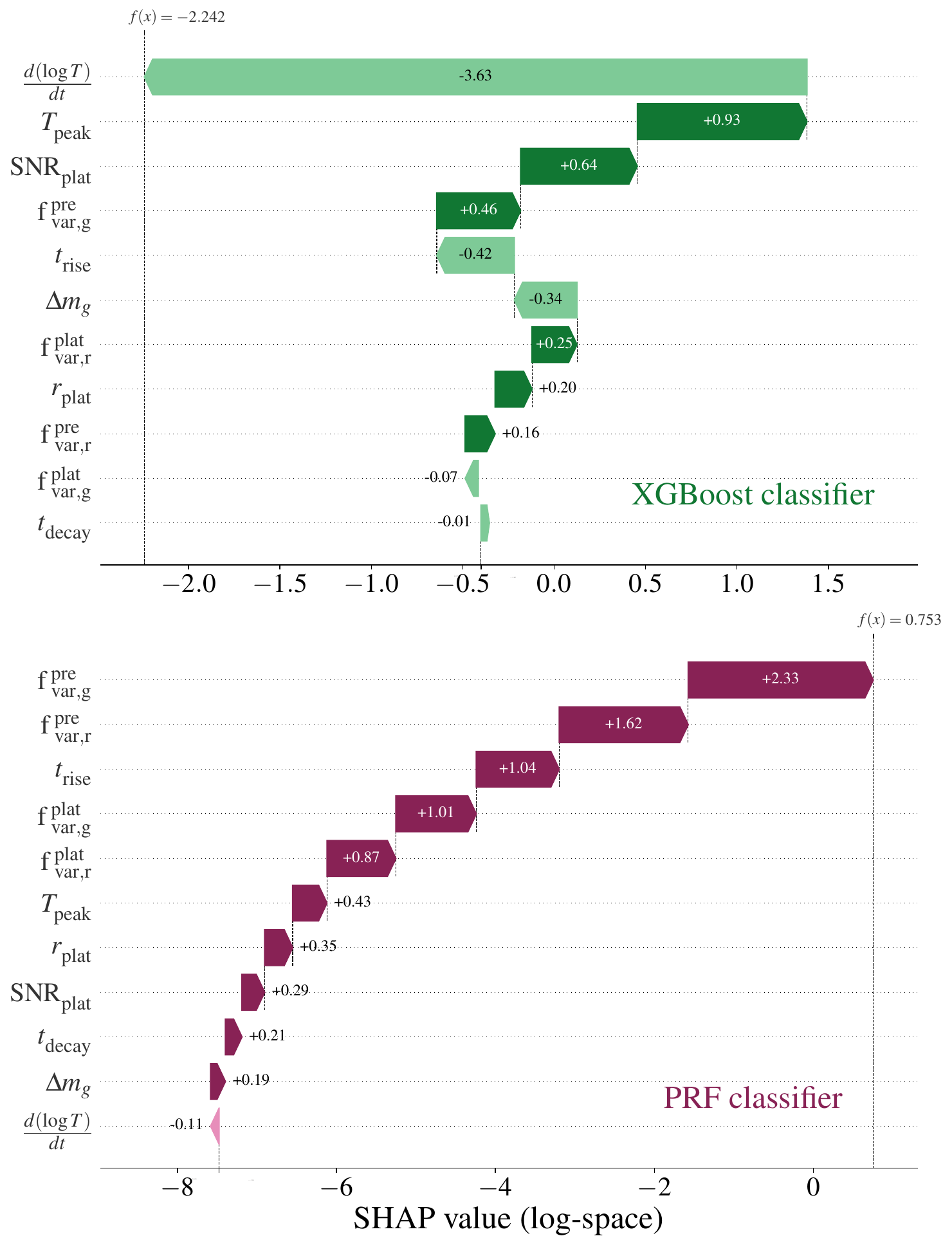}
              \caption{}
              \label{fig:prf_xgb_shap_comp_TDE2022hvp}
          \end{subfigure}
      \end{minipage}
      \caption{\red{ZTF lightcurves and SHAP waterfall plots for the two case studies, TDE2022emf in the left column and TDE2022hvp in the right
  column.} Panels (a) and (c) show the ZTF lightcurves, where thicker lines represent the median estimates of the fit and thinner lines represent random
  draws from the posterior distribution generated via MCMC sampling. Panels (b) and (d) show the corresponding SHAP waterfall plots, with the XGBoost
  classifier in the upper half and the PRF classifier in the lower half of each panel. Darker shades indicate a positive feature contribution, while
  lighter shades indicate a negative contribution. \red{Since both sources are spectroscopically confirmed TDEs, a well measured feature with values typical of the class is expected to contribute positively.}}
      \label{fig:shap_comp_cases}
  \end{figure*}

\red{Looking deeper into the feature level contributions, we consider two transients, TDE2022emf and TDE2022hvp, where a poorly constrained parameter is handled differently by the two classifiers (Figure \ref{fig:shap_comp_cases}). For TDE2022emf, } the sparse pre-peak data in Figure \ref{fig:PRFXGB_lcplot_TDE2022emf} yields a poorly constrained early-time model fit, with the measurement uncertainty reflected in the broad distribution of the MCMC posterior draws for the rise time parameter, $t_{\rm rise}$. The SHAP values in Figure \ref{fig:prf_xgb_shap_comp_TDE2022emf} detail how each algorithm processes this variance. XGBoost evaluates the median $t_{\rm rise}$ value as a rigid point estimate, ignoring the variance around the measurement. Because this specific median value deviates from a standard TDE (a very fast rise is more probable for a supernova), the gradient boosting framework penalizes the classification by assigning $t_{\rm rise}$ a strong negative contribution and pushing it to the top of the feature importance ranking. The PRF framework evaluates the same parameter using the probability density function derived from the MCMC errors. Integrating over the broad posterior distribution allows PRF to account for the probability mass aligning with standard TDE parameters, converting the parameter into a positive contributor. The algorithm simultaneously downweights the unconstrained parameter, dropping $t_{\rm rise}$ lower in the feature importance ranking to prevent the imprecise median from dominating the classification. 

\red{For TDE2022hvp, the lightcurve is well sampled very close to the peak (Figure \ref{fig:PRFXGB_lcplot_TDE2022hvp}), but poorly measured during the decay, particularly lacking proper multi-band coverage. This results in poorly constraining the temperature gradient parameter, which depends on the multi-band modeling, which drives the fit towards a steep temperature evolution, where the median $\log_{10} T$ reaches the upper limit of $10^6$ K before $t_{\rm plat}$, with the posterior draws spanning a wide range. The SHAP values in Figure \ref{fig:prf_xgb_shap_comp_TDE2022hvp} show the consequence. In the training sample, TDEs and AGN are characterized by temperature gradients close to zero, while supernovae extend towards negative values. The gradient measured here is large and positive, and XGBoost, taking the median at face value, assigns it a strong negative contribution at the top of the ranking, larger in magnitude than the combined positive contribution of all the remaining features. The source is consequently classified as a supernova with a probability of 0.98, leaving a TDE probability of only 0.009. Instead the PRF, using the broad posterior left by the sparse multi-band coverage during the decay, down-weights the same parameter to the least important feature. The classification is instead driven by the pre-transient variability features, which remain well measured, and the source is assigned a TDE probability of 0.68.}

This clearly shows how deterministic models could misclassify noisy data by forcing rigid point estimates on uncertain measurements, while the probabilistic framework absorbs this measurement uncertainty by distributing the impact across the probability density function. \red{For TDE2022emf, even though} XGBoost is affected by the exclusion of the uncertainties, the strong contribution of the remaining features ensures the final classification of this source remains a TDE across both models. \red{For TDE2022hvp, the remaining features are not sufficient, and the mis-measured temperature gradient is enough for XGBoost to classify the source as a non-TDE.} \red{These examples highlight}  the necessity of utilizing an uncertainty-aware, feature-based classifier that handles measurement variance directly, a critical requirement to prevent misclassifications when operating on sparse, low signal-to-noise data.

\subsection{Classifier performance at the faint end}\label{subsec:SNRcomp}

As evident from Figure \ref{fig:peakmag_unk}, the newly found TDE candidates mostly populate the faint end of the known population, aligning with the survey's detection limit. To better understand this behavior of the classifier near the detection limits of upcoming surveys, we test the performance of both models on a source with synthetically degraded photometry. To simulate these observations, the mean continuous flux for a given transient in individual filters is first evaluated at the original observation epochs. To estimate this mean flux, we employ a multi-band Gaussian Process framework incorporating a physically motivated mean function (Anilkumar et al., in prep). We select TDE2021axu as the reference TDE to calculate the mean flux profile for this experiment. We choose this source because the corresponding lightcurve has good coverage in both $g$ and $r$ bands with both classifiers classifying the source as TDE with high confidence. We do this to avoid any factors apart from the brightness and signal strength influencing the classification during the test. 

After determining the mean flux profile, we gradually degrade the peak strength by applying discrete dividing scaling factors, which corresponds to a decrease in peak flux. For each magnitude step, we execute 100 independent Monte Carlo iterations. During each iteration, we inject randomized photometric noise into the scaled light curve using an empirical, flux-dependent error model derived from the source's ZTF reference light curve. To construct this model, we first pool the flux and flux-uncertainty measurements across all photometric bands of the  observed light curve, yielding an empirical mapping between flux level and  photometric error. For a given perturbed flux value, we then draw an  uncertainty at random from all reference measurements whose fluxes fall within a fixed window of that value (defaulting to the nearest-neighbour measurement where the window is empty), thereby preserving the heteroscedastic, flux-dependent character of ZTF photometry rather than assuming a single global noise level. We apply this model point-by-point to the scaled Gaussian-process mean profile by assigning a sampled uncertainty to flux value and is then perturbed by drawing from a Gaussian centered on the scaled flux with a standard deviation equal to that sampled error. Repeating this procedure over the 100 iterations at each magnitude step produces an ensemble of realistic, progressively noisier light curves whose noise properties remain consistent with genuine ZTF observations at the corresponding flux level. 

\begin{figure}
    \centering
    \includegraphics[width=\linewidth, alt={Predicted TDE probability versus fading peak magnitude; PRF declines gradually to 0.45 while XGBoost collapses to near zero past magnitude 21.}]{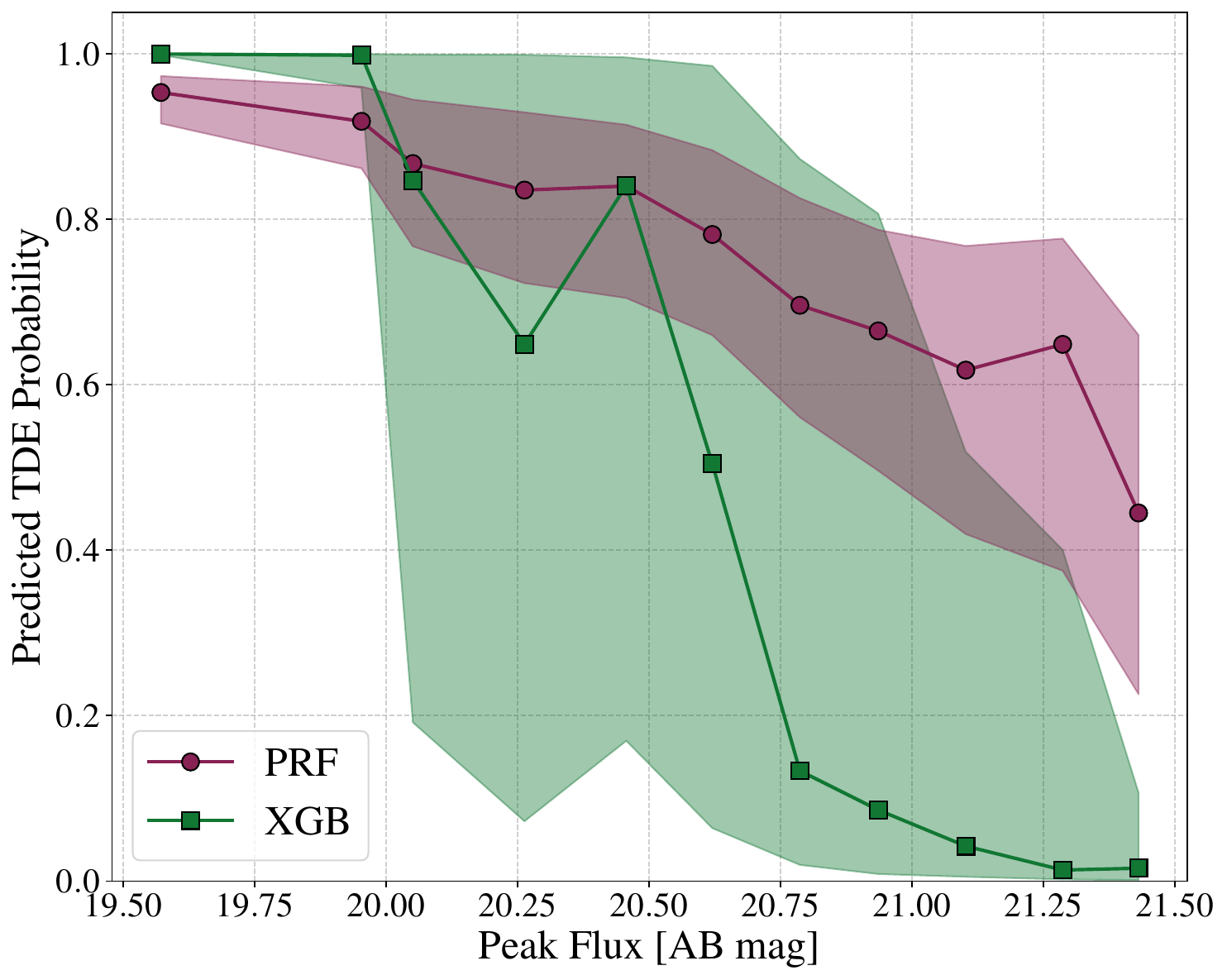}
    \caption{Classification robustness of the models under simulated signal-to-noise degradation. The x-axis represents the systematically reduced peak flux of the reference transient TDE2021axu, mapped to apparent AB magnitude. The markers indicate the median predicted TDE probability across 100 randomized photometric noise iterations at each magnitude bin, and the shaded regions represent the corresponding IQR for the PRF and XGBoost classifiers.}
    \label{fig:SNRvar}
\end{figure}

Following the calculation of the synthetic light curves, the parameter inference pipeline (Section \ref{subsec:featextract}) extracts new feature posteriors from every realization, which are subsequently evaluated by both classifiers. Figure \ref{fig:SNRvar} displays the predicted TDE probability as a function of decreasing peak flux. The distribution of the predicted TDE probabilities across the 100 iterations at each magnitude bin defines the median and the inter-quartile range, capturing the classification variance induced by the randomized photometric noise. The XGBoost model maintains a high classification probability until the variance of the injected randomized noise becomes prominent, past which the predicted probability drops rapidly from \red{84\% to 10\%} of the original value within \red{0.5} mag.

On the other hand, the PRF classifier exhibits a gradual reduction in predicted probability. As peak flux decreases and the relative variance of the injected measurement noise increases, the MCMC sampling produces broader posterior distributions for the extracted light curve parameters. The PRF architecture incorporates these expanding feature uncertainties, yielding a monotonic decay in the median TDE probability and accurately widening the inter-quartile range of the predictions in response to the reduced peak flux. The PRF output directly reflects the physical ambiguity inherent in low signal-to-noise data, demonstrating a greater sensitivity to data quality than the binary nature of XGBoost decisions. While based on a single source, this test illustrates that uncertainty-aware classification methods are better equipped to handle the low-flux regime expected from the Rubin Observatory, where the majority of transient alerts will lack the signal strength needed for robust deterministic classification.

\section{Conclusion and Future Directions}\label{sec:conclusion}

In this work, we presented a host-agnostic, uncertainty-aware classification framework for TDEs from the ZTF nuclear transient stream, and evaluated it against a deterministic XGBoost baseline. The main conclusions are as follows.

\begin{itemize}


\item[$\bullet$] The PRF prioritizes stable, low-variance features that generalize robustly across varying data quality, while XGBoost leverages physically motivated point estimates such as peak blackbody temperature and temperature gradient but remain blind to measurement uncertainty.

\item[$\bullet$] The PRF provides significantly higher classification stability for ambiguous or faint candidates, returning consistent probability distributions that broaden honestly with data quality, whereas XGBoost predictions for the same sources scatter erratically depending on training set composition.

\item[$\bullet$] While XGBoost achieves marginally higher recall in the high-precision scenario, the PRF rejects approximately 35\% more false positives in the high-recall scenario, demonstrating a substantially more efficient and conservative approach when completeness is prioritized.

\item[$\bullet$] The signal degradation test (Fig.~\ref{fig:SNRvar}) demonstrates that PRF degrades gradually as photometric quality decreases, maintaining a monotonic and widening probability distribution that reflects data quality, whereas XGBoost exhibits an abrupt collapse in predicted probability over a narrow magnitude range, making the PRF better suited for the low-flux regime.

\item[$\bullet$] Applying the trained models to the archival ZTF data, we identified 11 new TDE candidates from previously unclassified nuclear transients. We also flagged 1 classified supernova and 2 AGN which are potential TDE misclassifications.

\item[$\bullet$] Taken together, the comparison does not establish PRF as a strict replacement for deterministic classifiers, but rather positions the two architectures as complementary. XGBoost remains effective when features are well-constrained, while the PRF's value emerges in the low-SNR and sparsely-sampled regime by propagating measurement uncertainty into both the predicted probabilities and the feature-level contributions.

\end{itemize}

This work establishes a foundation for uncertainty-aware, host-agnostic TDE classification, but several natural directions remain open for future development to prepare for Rubin. The pre-transient and plateau variability features require a  baseline of observations prior to and following the transient event to be reliably estimated. The cadence and observing strategy of the Rubin Observatory may not always provide this baseline at sufficient depth, particularly for fainter sources at higher redshifts, which could degrade the informativeness of these features for a significant fraction of candidates. Additionally, the analytical lightcurve model employed in this work for feature extraction assumes a well-sampled transient profile. In the sparse data regime expected from Rubin, such analytical fitting will become increasingly unreliable. Moving toward, Gaussian Process-based feature extraction frameworks, which are better suited to irregularly sampled and sparsely observed lightcurves, will be an important step to make this pipeline functional in the Rubin era.

\red{One limitation of the present framework concerns the population it is applied to rather than the classifier itself. Candidates are drawn from a nuclear alert filter, so TDEs with a significant offset from the center of their host are excluded before the classifier is reached, and \citet{lanzaEarlyIdentificationOptical2026} make a related point in retaining a classifier variant without the host offset in order to stay sensitive to such events.  A second bias enters through the features themselves. The pre-transient fractional rms features respond to variability at the source position, so TDEs occurring in persistently variable nuclei are harder to separate from the AGN class, although the two candidates recovered from the labeled AGN sample in Section \ref{subsec:newTDEsearch} indicate that this is not prohibitive. Since every feature is measured on the transient lightcurve alone, relaxing the nuclear filter would be a natural way to extend the framework to the off-nuclear population.}

A further limitation concerns the comparison between the PRF and XGBoost, which are architecturally very different classifiers. XGBoost was selected as the baseline because it represents the most widely adopted architecture in the TDE classification literature, making the comparison practically meaningful. However, a more powerful direction would be to incorporate measurement uncertainty directly into a gradient boosting framework, combining the sequential residual-correction strength of boosting with the probabilistic feature handling of the PRF.

As the volume and diversity of transient alerts from the Rubin Observatory grows by orders of magnitude, when compared to existing surveys, the ability to propagate measurement uncertainty through the classification process will become essential for reliable TDE identification. This work demonstrates that uncertainty-aware, host-agnostic frameworks are a viable and necessary step toward that goal.

\section{Data Availability}\label{sec:dataavail}

The ZTF forced photometry underlying this work is publicly available on \href{https://doi.org/10.5281/zenodo.21628415}{Zenodo}\footnote{https://doi.org/10.5281/zenodo.21628415}. It comprises the baseline and extinction corrected light curves in $g$, $r$ and $i$ bands for all 12251 nuclear transient candidates, together with two catalog tables listing positions, extinction coefficients, and spectroscopic classifications with their source references. We also release the per-source PRF and XGBoost class probabilities for the unclassified sources, so that future searches for nuclear transients in ZTF can be compared directly against the rankings presented here.



\bibliographystyle{mnras}
\bibliography{biblo} 


\appendix
\onecolumn

\section{List of Tidal Disruption Events}
\begin{center}
    \scriptsize
    \renewcommand{\arraystretch}{1.15}
    \begin{tabular}{c|m{14em}|c|c|m{20em}}
        Plot ID& Name & Redshift & Peak magnitude ($r$ band) & Source \\
        \hline \hline
        \vspace{0.2em}
        1 & ZTF21aabiipy / TDE2021lo & 0.152 & 19.22  & \citet{2022TNSCR.620....1Y} \\
        2 & ZTF20achpcvt / TDE2020vwl & 0.0325 & 17.22  & \citet{2023MNRAS.522.5084G,yaoTidalDisruptionEvent2023} \\
        3 & ZTF21aapvvtb / TDE2021gje & 0.358 & 19.86  & \citet{2021TNSCR1723....1H} \\
        4 & ZTF21aaaokyp / TDE2021axu & 0.190 & 18.97  & \citet{yaoTidalDisruptionEvent2023} \\
        5 & ZTF22aabimec / TDE2022csn & 0.148 & 18.37  & \citet{Arcavi2022} \\
        6 & ZTF20abnorit / AT2020ysg & 0.277 & 18.65  & \citet{2023ApJ...942....9H} \\
        7 & ZTF19aapreis / TDE2019dsg & 0.0512 & 18.05  & \citet{vanVelzen21} \\
        8 & ZTF19aakiwze / AT2019cho & 0.193 & 19.6  & \citet{vanVelzen21} \\
        9 & ZTF21abmwftm / TDE2021uvz & 0.176 & 19.75  & \citet{2022TNSCR.925....1Y} \\
        10 & ZTF19aakswrb / TDE2019bhf & 0.1206 & 18.76  & \citet{vanVelzen21} \\
        11 & ZTF22abkfhua / TDE2022wtn & 0.0491 & 18.16  & \citet{2022TNSTR2885....1F} \\
        12 & ZTF20abjwvae / AT2020opy & 0.159 & 19.11  & \citet{2023ApJ...942....9H} \\
        13 & ZTF21acafvhf / TDE2021yte & 0.053 & 18.88  & \citet{yaoTidalDisruptionEvent2023} \\
        14 & ZTF22aaaedas / TDE2022rz & 0.107 & 19.04  & \citet{2022TNSCR.891....1H} \\
        15 & ZTF20abowque / AT2020qhs & 0.345 & 19.19  & \citet{2023ApJ...942....9H} \\
        16 & ZTF21abqtckk / TDE2021utq & 0.127 & 18.75  & \citet{yaoTidalDisruptionEvent2023} \\
        17 & ZTF22aaabovl / TDE2022aee & 0.124 & 18.15  & \citet{Yao2022} \\
        18 & ZTF19abhhjcc / AT2019meg & 0.152 & 19.38  & \citet{vanVelzen21} \\
        19 & ZTF20abfcszi / TDE2020mot & 0.070 & 18.28  & \citet{2023ApJ...942....9H} \\
        20 & ZTF19aabbnzo / TDE2018lna & 0.091 & 18.91  & \citet{vanVelzen21} \\
        21 & ZTF21abcgnqn / TDE2021nwa & 0.047 & 18.69  & \citet{2021TNSCR2155....1Y} \\
        22 & ZTF20acyydkh / TDE2021ack & 0.133 & 19.51  & \citet{2021TNSCR.732....1H} \\
        23 & ZTF20acnznms / TDE2020yue & 0.2042 & 18.54  & \citet{yaoTidalDisruptionEvent2023} \\
        24 & ZTF22aaahtqz / TDE2022bdw & 0.0378 & 17.58  & \citet{2022TNSCR.511....1A} \\
        25 & ZTF21aaeoitd / TDE2021jsg & 0.126 & 19.77  & \citet{2021TNSCR1221....1Y} \\
        26 & ZTF18abxftqm / TDE2018hco & 0.088 & 18.37  & \citet{vanVelzen21} \\
        27 & ZTF17aaazdba / TDE2019azh & 0.0222 & 15.35  & \citet{2021MNRAS.500.1673H,2022ApJ...925...67L,vanVelzen21} \\
        28 & ZTF21abxngcz / TDE2021yzv & 0.286 & 19.17  & \citet{yaoTidalDisruptionEvent2023} \\
        29 & ZTF20acvezvs / AT2020abri & 0.178 & 19.13  & \citet{yaoTidalDisruptionEvent2023} \\
        30 & ZTF20aamqmfk / AT2020ddv & 0.160 & 19.74  & \citet{2023ApJ...942....9H} \\
        31 & ZTF22aavvqyh / TDE2022pna & 0.095 & 18.59  & \citet{2022TNSCR2892....1Y} \\
        32 & ZTF19aarioci / TDE2019ehz & 0.074 & 18.27  & \citet{vanVelzen21} \\
        33 & ZTF19acspeuw / TDE2019vcb & 0.088 & 18.84  & \citet{2023ApJ...942....9H} \\
        34 & ZTF20abefeab / AT2020mbq & 0.093 & 19.02  & \citet{2023ApJ...942....9H} \\
        35 & ZTF22aaabqko / TDE2022emf & 0.081 & 19.65  & \citet{Hall2025} \\
        36 & ZTF22abajudi / TDE2022lri & 0.0328 & 17.97  & \citet{2024ApJ...976...34Y} \\
        37 & ZTF18actaqdw / AT2018lni & 0.138 & 19.49  & \citet{vanVelzen21} \\
        38 & ZTF18acaqdaa / TDE2018iih & 0.212 & 18.84  & \citet{vanVelzen21} \\
        39 & ZTF19aatylnl / AT2019eve & 0.0813 & 19.12  & \citet{vanVelzen21} \\
        40 & ZTF20acitpfz / TDE2020wey & 0.02741 & 18.12  & \citet{2023ApJ...942....9H} \\
        41 & ZTF21abjrysr / TDE2021sdu & 0.059 & 18.72  & \citet{yaoTidalDisruptionEvent2023} \\
        42 & ZTF22aagvrlq / TDE2022ibq & 0.395 & 18.98  & \citet{Yao2022ibq} \\
        43 & ZTF22aagyuao / TDE2022hvp & 0.112 & 16.91  & \citet{2022TNSAN.106....1F} \\
        44 & ZTF20acqoiyt / TDE2020zso & 0.0565 & 17.81  & \citet{Wevers2022, 2023ApJ...942....9H} \\
        45 & ZTF18aakelin / AT2020ocn & 0.070 & 19.91  & \citet{2023ApJ...942....9H} \\
        46 & ZTF19abzrhgq / TDE2019qiz & 0.0151 & 15.85  & \citet{Hung2021, Patra2022, Wu2025} \\
        47 & ZTF19abhejal / AT2019mha & 0.148 & 19.52  & \citet{vanVelzen21} \\
        48 & ZTF19abidbya / TDE2019lwu & 0.117 & 18.97  & \citet{vanVelzen21} \\
        49 & ZTF22abegjtx / TDE2022upj & 0.052 & 18.26  & \citet{2024ApJ...977..258N} \\
        50 & ZTF22aacgcwv / TDE2022dyt & 0.072 & 18.65  & \citet{2022TNSCR1036....1S} \\
        51 & ZTF20abisysx / TDE2020nov & 0.084 & 17.69  & \citet{Frederick2020, Earl2025} \\
        52 & ZTF21aauuybx / TDE2021jjm & 0.153 & 18.96  & \citet{yaoTidalDisruptionEvent2023} \\
        53 & ZTF20abgwfek / TDE2020neh & 0.062 & 17.82  & \citet{2022NatAs...6.1452A} \\
        54 & ZTF20aahmtso / TDE2022gri & 0.028 & 18.01  & \citet{2022TNSAN..99....1Y} \\
        55 & ZTF22aaddwbo / TDE2022gdw & 0.105 & 19.4  & \citet{Hammerstein2022} \\
        56 & ZTF20aabqihu / TDE2020pj & 0.068 & 18.78  & \citet{2023ApJ...942....9H} \\
        57 & ZTF21abhrchb / TDE2021qth & 0.0805 & 18.53  & \citet{Yao2023} \\
        58 & ZTF19accmaxo / TDE2019teq & 0.0874 & 19.41  & \citet{2023ApJ...942....9H} \\
        59 & ZTF21aanxhjv / TDE2021ehb & 0.017 & 17.34  & \citet{2022ApJ...937....8Y} \\
        \hline
    \end{tabular}
    \captionof{table}{List of TDEs used in the paper as part of the training data.}
    \label{tab:tde_list}
\end{center}

\clearpage
\section{Lightcurves of the new TDE candidates}
\begin{center}
    \centering
    \includegraphics[width=0.98\textwidth, alt={Grid of 14 light curves of the high-precision TDE candidates, each with r and g photometry, occasional Swift/UVOT points, and a fitted rise-and-decay model.}]{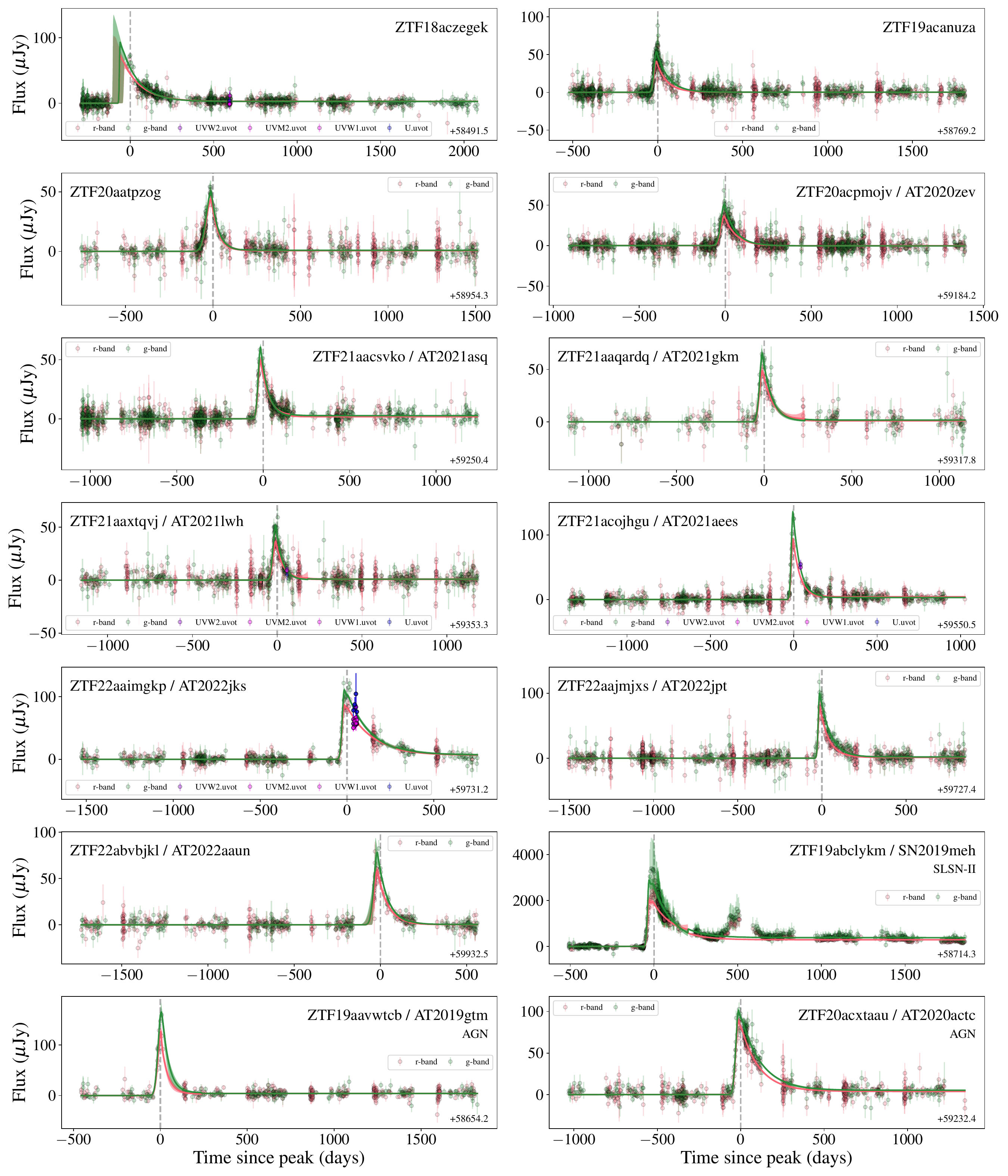}
    \captionof{figure}{The extinction-corrected lightcurves of the 14 TDE candidates from the high-precision search. The \textit{Swift} UVOT data \citep{Gehrels2004, Roming2005} are also plotted where available. The host-subtracted flux for the \textit{Swift} data is calculated using the procedure described in \citet{vanVelzen21}. The best-fit curve is shown for each transient, along with its 1$\sigma$ bands. If a transient is a classified source in our training set, the classification is given below the source name. The bottom-right corner lists the original peak time in MJD.}
    \label{figA:LCplots}
\end{center}

\bsp	
\label{lastpage}
\end{document}